\documentclass[journal]{IEEEtran}
\IEEEoverridecommandlockouts
\usepackage{fancyhdr}
\usepackage{graphicx}
\usepackage[english]{babel}  % No "Tabla" replacement

\usepackage[utf8]{inputenc}
\usepackage{color}
\usepackage{hyperref}
\usepackage{wrapfig}
\usepackage{array}
\usepackage{multirow}
\usepackage{adjustbox}
\usepackage{nccmath}
\usepackage{subfigure}
\usepackage{amsfonts,latexsym} % to have availability of various symbols
\usepackage{enumerate}
\usepackage{booktabs}
\usepackage{float}
\usepackage{threeparttable}
\usepackage{array,colortbl}
\usepackage{ifpdf}
\usepackage{rotating}
\usepackage{cite}
\usepackage{stfloats}
\usepackage{url}
\usepackage{listings}
\usepackage{physics}
\usepackage{longtable}
\usepackage{supertabular}
\usepackage[table, svgnames, dvipsnames]{xcolor}
\usepackage{lipsum}
\usepackage{logicpuzzle}
\usepackage{tikz}
\usetikzlibrary{shapes.geometric, arrows, positioning}
\usepackage{overpic}   % For overlaying text on figures
\usepackage[dvipsnames]{xcolor}

\tikzstyle{startstop} = [rectangle, rounded corners, minimum width=3cm, minimum height=1cm,text centered, draw=black, fill=red!30]
\tikzstyle{process} = [rectangle, minimum width=3cm, minimum height=1cm, text centered, draw=black, fill=blue!30]
\tikzstyle{decision} = [diamond, minimum width=3cm, minimum height=1cm, text centered, draw=black, fill=green!30]
\tikzstyle{arrow} = [thick,->,>=stealth]
\usepackage{multirow, makecell}
\usepackage{tabu}

\usepackage{booktabs} % for better table lines

\usetikzlibrary{shapes.geometric, arrows, positioning}

\usepackage{tikz}
\usetikzlibrary{shapes.geometric, arrows, positioning, decorations.pathmorphing}

\tikzstyle{block} = [rectangle, draw, rounded corners, text centered, minimum height=1.2cm, minimum width=2.5cm]
\tikzstyle{arrow} = [thick,->,>=stealth]
\tikzstyle{wavy} = [decorate, decoration={snake, amplitude=0.7mm, segment length=3mm}, draw]

\newcolumntype{P}[1]{>{\centering\arraybackslash}p{#1}}  %% A new type of column called P is created.

\renewcommand\IEEEkeywordsname{Keywords}
\graphicspath{ {Figs/} }  %%Ruta donde se encuentran las imágenes, que esté vacio indica que las imagenes están dentro de la misma carpeta que contiene el archivo .tex

\newcommand{\MYhead}{\smash{\scriptsize
\hfil\parbox[t][\height][t]{\textwidth}{\centering
\underline{\hspace{ \textwidth}}}\hfil\hbox{}}}
\makeatletter
\def\ps@headings{%
\def\@oddhead{\MYhead}%
\def\@evenhead{\MYhead}}%
\def\ps@IEEEtitlepagestyle{%
\def\@oddhead{\MYhead}%
\def\@evenhead{\MYhead}}%
\makeatother
\begin{document}
%%%%%%%%%%%%%%%%%%%%%%%%%%%%
%%% DOCUMENT TITLE %%%
%%%%%%%%%%%%%%%%%%%%%%%%%%%%
\title{Device-Independent Quantum Key Distribution: Protocols, Quantum Games, and Security}

%%%%%%%%%%%%%%%%%%%%%%%%%%%%
%%%%%%%%% AUTHORS %%%%%%%%%
%%%%%%%%%%%%%%%%%%%%%%%%%%%
% \author{Syed M. Arslan,~  Saif Al-Kuwari~, ~ M. T. Rahim, ~ Hashir Kuniyal}
        \author{
    \IEEEauthorblockN{Syed M. Arslan,~  Saif Al-Kuwari~, ~ M. T. Rahim, ~ Hashir Kuniyal}\\
    \IEEEauthorblockA{\IEEEauthorrefmark{1}Qatar Center for Quantum Computing, College of Science and Engineering, \\ Hamad Bin Khalifa University, Doha, Qatar
    \\{Corresponding Author:} syan54988@hbku.edu.qa}
    % \IEEEauthorblockN{}\\
    % \IEEEauthorblockA{\IEEEauthorrefmark{2}Institute of Physics, Faculty of Physics, Astronomy and Informatics, Nicolaus Copernicus University in Torun, ul. Grudziadzka 5, 87-100 Torun, Poland
    % \\{Corresponding Author:} hkuniyil@hbku.edu.qa}
}
				
				% stops a space
%\thanks{El presente documento corresponde al articulo final del proyecto de práctica de ingeniería electrónica 3 presentado en la Universidad Central durante el periodo 2017-1.}} %\thanks appends a footnote where you can put some information about the nature of the document.
%%%%%%%%%%%%%%%%%%%%%%%%%%%

% Command that indicates the generation of the title
\maketitle

%%%%%%%%%%%%%%%%%%%%%
%%%%%% Abstract %%%%%%
%%%%%%%%%%%%%%%%%%%%%
\begin{abstract}
  Quantum Key Distribution (QKD) is based on the laws of quantum mechanics to enable provably secure communication. Despite its theoretical security promise, practical QKD systems are vulnerable to serious attacks, including side‐channel attacks and detector loopholes, and assumes a trusted device characterization. 
Device‐Independent Quantum Key Distribution (DIQKD) overcomes these limitations by relying solely on observed nonlocal correlations, certified through Bell inequality violations, thereby removing assumptions about the internal workings of the measurement devices.

In this paper, we first review the foundational principles underlying DIQKD, including Bell tests and security definitions. We then examine a range of protocol designs, including CHSH-based schemes, and non-local game frameworks, alongside  with their security proofs. We also assess recent experimental implementations and discuss source architectures, detection technologies, and finite-key analyses. Finally, we identify current open problems, such as noise tolerance, generation rates, and integration with quantum networks, and outline promising directions for future research to realize robust, high-performance DIQKD.
\end{abstract}
% In the abstract it is not recommended to place bibliographic citations.

%%%%%%%%%%%%%%%%%%%%%%
%%% KEYWORDS %%%
%%%%%%%%%%%%%%%%%%%%%%
\begin{IEEEkeywords}
Quantum Key Distribution, Device Independent QKD, Secure Communication
\end{IEEEkeywords}
%%%%%%%%%%%%%%%%%%%%%%
%\IEEEpeerreviewmaketitle

%%%%%%%%%%%%%%%%%%%%%%%%%%%%%%%%%%%%%
%%% FIRST SECTION OF THE DOCUMENT %%%
%%%%%%%%%%%%%%%%%%%%%%%%%%%%%%%%%%%%%
\section{Introduction}
\IEEEPARstart{Q}{u}antum key distribution (QKD) is a promising solution to the imminent threat of quantum computers. With successful demonstrations on various platforms, the current focus of research in QKD is to enhance the robustness of its protocols against attacks. Although QKD is theoretically unconditionally secure, closing all practical loopholes is still an open problem. In practice, QKD systems exhibit several vulnerabilities, particularly due to nonideal detector behavior and unintended information leakage through side channels. The first quantum attack on a QKD system was demonstrated using a time-shift attack by Zhao \emph{et al.} \cite{zhao2008quantum}. This work spurred further research into practical vulnerabilities, leading to the demonstration of blinding attacks \cite{sauge2011controlling, gerhardt2011full, lydersen2010hacking}. These blinding attacks exploit a loophole in single-photon detectors by using bright light to force the avalanche photodiodes out of their sensitive Geiger mode and into a linear, classical regime where an eavesdropper can fully control detection outcomes. They have been demonstrated using continuous wave illumination \cite{gerhardt2011full}, short bright trigger pulses \cite{lydersen2010hacking}, and by inducing thermal effects that shift the detector bias characteristics \cite{lydersen2010thermal}. Over the years, researchers have identified a diverse range of attack strategies that exploit both the fundamental properties of quantum communication channels and the practical imperfections of the devices used. These strategies can be broadly categorized into two groups: main-channel attacks and side-channel attacks. Main-channel attacks focus directly on the quantum channel used for the transmission of qubits between the communicating parties. An example of a main-channel attack is the Photon Number Splitting (PNS) attack \cite{lutkenhaus2002quantum,pirandola2020advances} where Eve exploits the multi-photon pulses inadvertently generated by imperfect single-photon sources. By intercepting one photon while allowing the others to continue unimpeded, she can gain partial information about the key without introducing detectable disturbances in the quantum correlations. In contrast, side-channel attacks exploit vulnerabilities that arise from the physical implementation of QKD systems. These attacks take advantage of unintended information leakage from the hardware. For example, in a Trojan Horse attack \cite{gisin2006trojan}, Eve injects additional photons or signals into the devices. This can cause sensitive information, such as the basis settings used in the key generation process, to be inadvertently revealed. Such attacks can cause a major setback to the practical utility of QKD. To address these issues, Device-Independent Quantum Key Distribution (DIQKD) has emerged as a potential solution.

DIQKD methods are capable of achieving a loophole-free quantum communication network while reducing reliance on trusted devices (i.e., the nodes belonging to Alice and Bob). In addition, DIQKD decouples protocol details from the underlying infrastructure, effectively eliminating many quantum attack vectors that depend on specific device models. However, DIQKD also introduces several challenges. Experimentally, it requires significant resources and suffers from low efficiency. It also faces major issues in feasibility in scalability when implemented. Moreover, DIQKD relies on strong assumptions, such as the validity of Bell test violations, the absence of hidden correlations between devices, and the requirement that Alice and Bob’s measurement settings, that is, their choices of polarization or spin measurement bases, are generated independently and unpredictably, with no influence from the devices themselves or any eavesdropper \cite{zapatero2023advances}. These assumptions are considered strong because even slight deviations, e.g., due to imperfections in the entanglement source or subtle device correlations, could undermine the security guarantees of the system. These assumptions will be revised in section \ref{DIQKD} (DIQKD), where we will explain why they present major challenges to practical DIQKD implementations.

In this paper, we provide a comprehensive review of the state of DIQKD while covering theoretical foundations and real-world applications of DIQKD. In particular, we investigate key theoretical concepts underpinning DIQKD, explore notable protocols, and review the current experimental efforts that have brought theories into reality.

%\subsection*{Outline of the Paper}
The rest of this paper is organized as follows: Section \ref{QKD}  Section \ref{DIQKD} provides a mathematical background on DIQKD and discusses the practical implementation of different types of DIQKD systems. Section \ref{games} reviews different types of non-local games and how they are used for DIQKD. Section \ref{security} provides a comprehensive study of the potential attacks that DIQKD can experience and research works that minimize the effect of these attacks. Section \ref{open} provides a discussion on the potential open problems for future research. Finally, section \ref{conclusion} concludes the survey. 
%%%%%%%%%%%%%%%%%%%%%%%%%%%%%%%%%%%%%
%%%%% THEORETICAL FRAMEWORK SECTIONS %%%%
%%%%%%%%%%%%%%%%%%%%%%%%%%%%%%%%%%%%%

%%%%%%%%%%%%%%%%%%%%%%%%%%%%%%
%%%%% CITAR BIBLIOGRAFIA %%%%%
%%%%%%%%%%%%%%%%%%%%%%%%%%%%%%
\section{Quantum Key Distribution (QKD)} \label{QKD}
Unlike traditional cryptographic methods that rely on hard mathematical problems for security, such as finding prime factors of large numbers or solving the discrete log problem, QKD is a new method to exchange encryption keys based on the laws of quantum mechanics. This makes it more secure against future technologies, including quantum computers, which can provably break today's cryptographic systems with access to sufficient quantum resources. QKD ensures that any attempt to eavesdrop on the communication is detectable. This is possible due to the unique quantum properties, including the \emph{no-cloning theorem}, which prevents copying of quantum information, and \emph{Heisenberg’s uncertainty principle}, which ensures that any measurements will inevitably disturb the system. 

\begin{figure*}[t]  % 't' places it at the top of the page
    \centering
    \begin{tikzpicture}
        % Include the PDF figure
        \node[anchor=south west, inner sep=0] (image) at (0,0)
            {\includegraphics[width=0.95\textwidth]{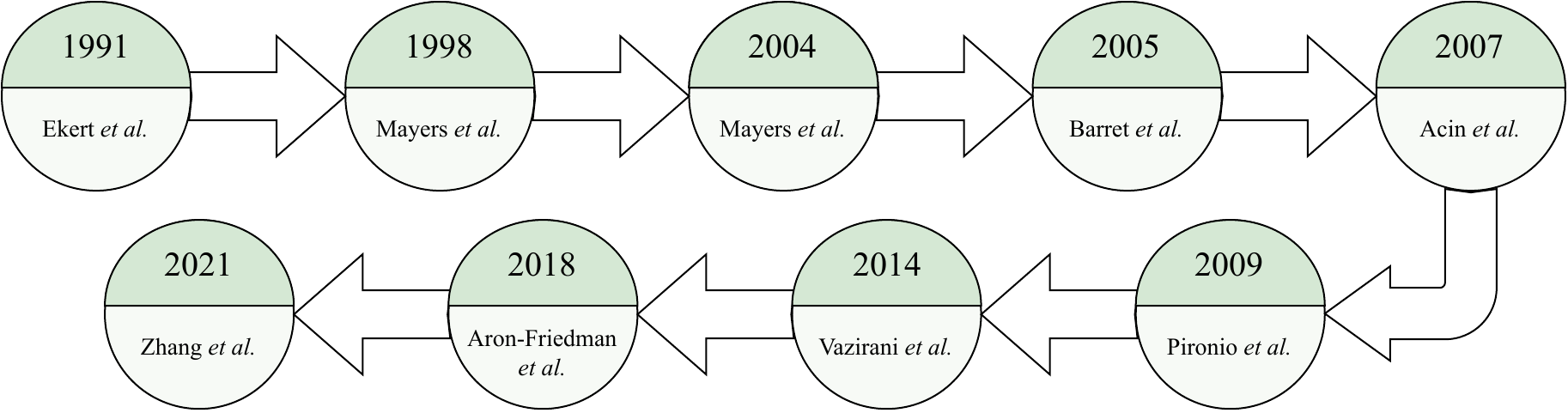}};
        
        % Overlay the text at specific locations
        \node[anchor=north west, font=\footnotesize] at (-0.5,5.5) 
            {\shortstack{First entanglement-based \\ QKD protocol \cite{ekert1991quantum}}};
        \node[anchor=north west, font=\footnotesize] at (3,5.5) 
            {\shortstack{‘Self-checking’ source \\ using EPR-Bell theorem \cite{mayers1998quantum}}};
        \node[anchor=north west, font=\footnotesize] at (7.7,5.5) 
            {\shortstack{Self-testing of \\ devices \cite{mayers2003self}}};
        \node[anchor=north west, font=\footnotesize] at (10.5,5.7) 
            {\shortstack{First quantitative link \\ between Bell violations and \\ key generations \cite{barrett2005no}}};
        \node[anchor=north west, font=\footnotesize] at (14.7,5.7) 
            {\shortstack{Security of robust \\ DI-QKD under collective \\ attacks \cite{acin2007device}}};
        \node[anchor=south west, font=\footnotesize] at (0.65,-1.25) 
            {\shortstack{Successful DI-QKD \\ demonstration over 400 \\ meters \cite{zhang2022device}}};
        \node[anchor=south west, font=\footnotesize] at (4,-1) 
            {\shortstack{Entropy Accumulation Theorem \\ (EAT) \cite{arnon2019simple}}};
         \node[anchor=south west, font=\footnotesize] at (8.3,-1) 
            {\shortstack{Complete security proof \\ of DI-QKD\cite{vazirani2014fully}}};
         \node[anchor=south west, font=\footnotesize] at (12.35,-1.25) 
            {\shortstack{Security proof of \\ DI-QKD against \\ general attacks \cite{pironio2009device}}};
    \end{tikzpicture}

    \caption{Key theoretical and experimental advancements in DIQKD.}
    \label{fig:widefig}
\end{figure*}
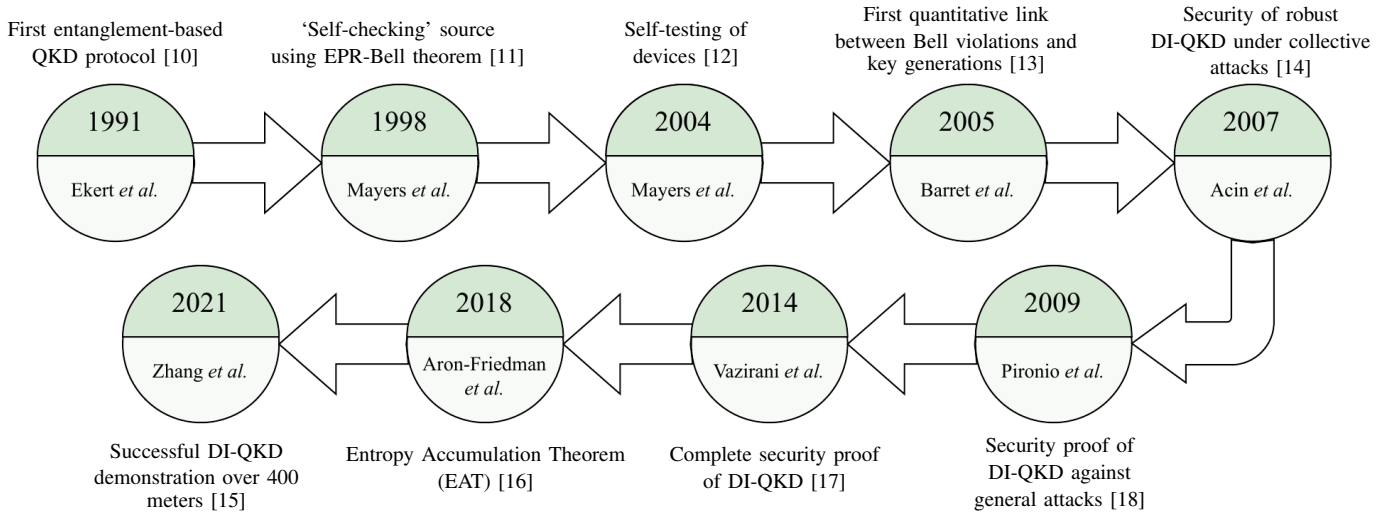

\subsection{Protocols}
QKD schemes can be broadly divided into two main categories based on the type of quantum states used:

\textbf{Discrete-Variable (DV) QKD:} DV-QKD protocols encode information into individual quanta (typically single photons) using discrete degrees of freedom, such as polarization, phase, or time bins. A classic example is the BB84 protocol \cite{bennet1984quantum}, which was the first to demonstrate the principles of quantum cryptography. Other protocols, such as E91 \cite{ekert1991quantum} and BBM92 \cite{bennett1992phys}, make use of quantum entanglement to establish secure keys. The security of these protocols rests on fundamental quantum principles like the no-cloning theorem and the disturbance caused by measurement. Further developments in DV-QKD include the decoy-state method \cite{lo2005decoy}, which improves security when using weak coherent pulses by countering photon number splitting attacks.

In contrast to DV-QKD, \textbf{Continuous-Variable (CV) QKD} encodes information in the continuous degrees of freedom of light, such as the amplitude and phase quadratures. This approach can leverage standard telecommunication components such as homodyne or heterodyne detectors, potentially enabling higher key rates and easier integration into existing optical networks. One influential protocol in this domain was proposed by Grosshans and Grangier (2002) \cite{grosshans2002continuous}, where coherent states are used in conjunction with homodyne detection. A comprehensive review of CV-QKD, including its theoretical framework and experimental progress, is provided in Weedbrook et al. (2012) \cite{weedbrook2012gaussian}.

On the other hand, the approach to state distribution categorizes protocols into two types: 
\begin{itemize}
    \item \emph{Prepare-and-Measure Schemes:} In these protocols, one party (traditionally Alice) prepares the quantum states and sends them to the other party (Bob), who then performs measurements on the received states. BB84 is the most well-known example of this approach. Prepare-and-measure schemes are common in both DV and CV implementations. For instance, DV protocols like BB84 use polarization states, while many CV protocols rely on preparing coherent states that are subsequently measured.
 \item \emph{Entanglement-Based Schemes:} Here, both Alice and Bob share parts of an entangled quantum state. The security of the key is inferred from the correlations between their measurement outcomes. E91 and BBM92 are prominent examples of entanglement-based DV-QKD. Moreover, entanglement-based approaches are not exclusive to DV protocols; CV entanglement-based protocols also exist \cite{djordjevic2024entanglement}, though they often present different technical challenges and advantages.
\end{itemize}
QKD protocols, whether they use discrete or continuous variables and whether they employ prepare-and-measure or entanglement-based approaches, rely on several key assumptions to ensure their security. These assumptions form the backbone of the theoretical guarantees of the protocols, and any deviation from them can undermine the overall security. Traditional QKD protocols, such as BB84 \cite{bennet1984quantum}, E91 \cite{ekert1991quantum}, and BBM92 \cite{PhysRevLett.68.557} rely on trusted quantum devices. To function correctly, the following assumptions are made:
\begin{enumerate}
    \item Quantum Mechanics: The protocols assume that quantum mechanics is correct. A critical property is that nonorthogonal quantum states that are used to encode information cannot be perfectly distinguished. This creates an inherent trade-off: any attempt by an eavesdropper to extract information will inevitably introduce disturbances, which underpin the security proofs.
    \item Secure Laboratories: It is assumed that the laboratories of the legitimate users (Alice and Bob) are completely secure, that there is no leakage of information in or out without their explicit approval. Since all secret key processing occurs within these labs, any leakage would jeopardize key secrecy. Although maintaining such secure environments is challenging, practical measures like tamper-resistant casings and strict access controls can help enforce this assumption. 
    \item Trustworthy Implementation: The implementation of QKD devices is expected to behave exactly as modeled in the security proofs. This "double trust" spans both the internal behavior of the devices and their consistency with theoretical models.
\end{enumerate}
However, on the last assumption, successful attacks have demonstrated that this assumption can be vulnerable, necessitating the development of countermeasures and certification procedures. One countermeasure is Device-Independent QKD, which seeks to relax the need to trust the hardware.
Additionally, randomness generation is crucial in QKD. While many conventional systems rely on pseudorandom number generators (PRNGs), QKD demands true randomness to uphold its information-theoretic security. This is typically achieved using true random number generators (TRNGs) or quantum random number generators (QRNGs), which derive unpredictability from physical or quantum processes rather than deterministic algorithms.

\subsection{Attacks}
\noindent Although the assumptions QKD makes provide a strong foundation for QKD security, Eve (eavesdropper) can exploit them to perform several quantum attacks. These attacks can generally be classified as main channel attacks and side channel attacks.

\paragraph{Main-channel attacks.} These attacks target the primary quantum communication channel between Alice and Bob, where qubits are transmitted. Examples of these attacks include
\begin{itemize}
    \item \textit{Photon Number Splitting (PNS) Attack \cite{lutkenhaus2002quantum,pirandola2020advances}:} PNS attacks take advantage of the weak coherence in single-photon sources and imperfections in photon detectors. By exploiting the possibility of multi-photon pulses, the attacker can intercept and measure a photon without disturbing the overall quantum correlations significantly. As a result, the attack can go undetected as it does not introduce noticeable errors in the quantum transmission.
    \item \textit{Intercept-and-Resend Attack:} In this attack, Eve intercepts each qubit sent from Alice to Bob, measures it in a chosen basis, and then prepares and forwards a new qubit based on her measurement outcome. Because she cannot perfectly clone an unknown quantum state, this process disturbs the transmitted qubits, degrading the correlations between Alice and Bob’s measurements and raising the observed error rate.
    
%     Eve intercepts the qubits being transmitted between Alice and Bob and resends new qubits based on her measurement results. Because of the quantum no-cloning and measurement disturbance principles, this introduces errors that Alice and Bob can later detect.
% %
    %\item \textit{Entanglement-Based Attack:} Eve introduces an ancillary qubit entangled with Alice’s or Bob’s qubits, allowing her to extract information without direct interception.
%
    \item \textit{Man-in-the-Middle Attack\cite{amellal2023quantum}:} Although less common when an authenticated classical channel is available, if the classical channel used for basis reconciliation is not properly authenticated, an adversary can position herself between Alice and Bob. She then establishes separate keys for each, effectively controlling the communication. (Note that MITM can target both the quantum and classical channels, but here we focus on its effect on the overall key-exchange process.)

    \item \textit{Channel Tampering Attacks}\cite{ghosh2017different}\textit{:}
    An attacker can physically disrupt or alter the quantum channel (for example, by cutting or jamming the fiber) to prevent key generation. Although this attack does not extract key information, it does compromise the availability of the secure channel.
\end{itemize}
\paragraph{Side-Channel attacks.} These attacks exploit unintended information leakage from QKD devices due to their inherent imperfections.
\begin{itemize}
    \item \textit{Trojan Horse Attack \cite{gisin2006trojan}:} In this attack, the adversary injects additional photons or signals into the quantum communication devices, such as Alice's source or Bob's detector. The attack exploits any unintended leakage or interactions between the quantum system and its environment, allowing the attacker to extract sensitive information about the quantum communication without directly interfering with the transmitted quantum states.
    \item \textit{Detector Blinding Attack \cite{lydersen2010hacking}:} In a detector blinding attack, the attacker sends strong optical signals to overwhelm the detectors in a QKD system, causing them to malfunction, producing incorrect readings of the quantum states. This manipulation does not involve tampering with the quantum states but rather exploits imperfections in the detector components, leading to faulty outcomes in the key exchange process.  
    \item \textit{Electromagnetic side-channel attack} \cite{durak2021attack,baliuka2023deep}\textit{:}
    This attack arises when sensitive components in a QKD setup, such as detectors, random number generators, or high-speed electronics, unintentionally emit radio-frequency (RF) or electromagnetic signals correlated with their internal states. An adversary monitoring these signals can infer crucial details of the QKD process, such as basis settings or detection events, thereby undermining the protocol's security.
    \item \textit{Phase-Remapping Attack} \cite{fung2007phase} \textit{:}
    In some QKD systems, the phase modulators used to encode quantum states can be imperfect. In a phase-remapping attack, Eve exploits these imperfections by sending signals that cause the modulator to output a phase different from the intended value. By remapping the phases, Eve can correlate her measurements with the actual key without introducing noticeable errors.
    \item \textit{Nonrandom-Phase Attack} \cite{sun2012partially} \textit{:}
    QKD protocols assume that the phase modulation is truly random. However, patterns may emerge if the phase modulator has biases or uses a pseudo-random number generator. In a nonrandom-phase attack, Eve exploits these predictable patterns to guess the encoded information, thereby compromising the security of the protocol.
    \item \textit{Double-Click Attack} \cite{zhang2011feasibility} \textit{:}
    In some QKD implementations, when both detectors register a click (a ``double-click”), a random bit value is assigned. Eve can exploit this behavior by inducing double clicks through tailored attack pulses. By doing so, she may introduce a bias or gain partial information about the key without significantly increasing the overall error rate.
    \item \textit{Fake-State Attack} \cite{makarov2007faked} \textit{:}
    Eve prepares and sends fake quantum states that mimic the expected behavior of genuine signals. She can force Bob’s detectors to register predetermined outcomes by carefully designing these states. This control over the detection outcomes allows Eve to extract key information while keeping the error rate within tolerable limits for Alice and Bob.
    \item \textit{Time-Shift Attack} \cite{zhao2008quantum} \textit{:}
    Due to slight timing differences and efficiency variations among Bob’s detectors, Eve can delay or advance the arrival time of the qubits so that one detector is more likely to click than the others. This selective triggering leaks partial information about the key.
    \item \textit{ detector-after-gate attack} \cite{wiechers2011after} \textit{:}
    In a detector-after-gate attack, Eve takes advantage of the brief recovery period immediately after a gated avalanche photodiode’s active (``on”) window. Although the gate has closed, the diode hasn’t yet fully quenched and remains partially sensitive for a few nanoseconds, registering photons via residual bias or after-pulsing. During this time, Eve sends carefully timed light pulses to trigger clicks. This lets her bias the detection outcomes without causing large error rates, potentially leaking key information.

\end{itemize}

\subsection{Countermeasures}\label{counter}
Traditional QKD protocols are based on trusted device models and a suite of predefined countermeasures to resist various attacks. For example, the security vulnerabilities exploited in the Photon Number Splitting (PNS) attack are mitigated using decoy-state protocols, as demonstrated in Lo, Ma, \& Chen (2005)\cite{lo2005decoy}; Hwang \emph{et al.} (2005) \cite{hwang2003quantum}. Intercept-and-resend attacks are typically detected through analysis of the quantum bit error rate (QBER) and countered by robust privacy amplification procedures, as detailed in Bennett \& Brassard (1984) \cite{bennett1984proc}; Shor \& Preskill (2000) \cite{shor2000simple}. Additionally, man-in-the-middle (MITM) attacks are prevented by using authenticated classical channels, following the approaches described in Maurer (1997) \cite{maurer1997information}.

On the side-channel front, several hardware-based strategies are employed. Trojan Horse attacks are blocked with optical isolators and wavelength filters, as discussed in Gisin \emph{et al.} (2006) \cite{gisin2006trojan}. Temporal gating is implemented to ensure that detectors respond only during predetermined time windows, a method that has been reviewed in Restelli \emph{et al.} (2010) \cite{restelli2010improved}. Detector blinding attacks are countered by adopting measurement device-independent QKD (MDI-QKD) protocols Lo \emph{et al.} (2012) \cite{lo2012measurement}, which effectively shift the security trust from detectors to verification of entanglement between communicating parties. Furthermore, electromagnetic side-channel attacks are mitigated by employing physical layer shielding and filtering techniques to suppress unintended RF and electromagnetic emissions. Durak et al. (2021) \cite{durak2021attack}, Baliuka et al. (2023) \cite{baliuka2023deep}.

\subsection{Challenges}
Despite the promising countermeasures in Section \ref{counter} that have been developed to thwart known QKD attacks, practical QKD implementations face several challenges that can undermine these solutions. Real-world systems rarely achieve the idealized behavior assumed in security proofs, leading to vulnerabilities even when countermeasures are in place. Some of the key practical challenges facing QKD include:
\begin{itemize}
    \item Imperfect Single-Photon Sources: Most QKD systems rely on weak coherent pulses rather than true single-photon sources. While decoy-state protocols help mitigate Photon Number Splitting (PNS) attacks, they require precise control over pulse intensity and phase. Any fluctuation or miscalibration can inadvertently allow multi-photon emission, providing an opening for an eavesdropper \cite{lo2005decoy}.
    \item Non-Ideal Detector Behavior:
    Detectors are assumed to have uniform efficiency and precise timing characteristics, but in practice, they suffer from efficiency mismatches, temporal jitter, and other imperfections. Even with temporal gating, where detectors are active only during narrow time windows, subtle differences in detector responses can be exploited by time-shift or faked-state attacks \cite{qi2005time}. For example, in the E91 protocol, Alice prepares a qubit on a basis of her choice (Hadamard or standard), while Bob measures the qubit using a basis rotated by $\pi/8$. In a realistic scenario, if Bob’s measurement device has flaws or miscalibration, these can be misinterpreted as noise. However, an eavesdropper like Eve can exploit these imperfections through a detector blinding attack where using a high-intensity laser would allow Eve to force Bob's detector into a different operational mode (e.g., from Geiger to linear mode), thereby compromising the security of the protocol without triggering noticeable anomalies. %This example shows that even small imperfections in detectors can open the door to significant security risks.
    \item Uncharacterized Side Channels:
    Many countermeasures assume complete and accurate characterization of the QKD devices. However, practical systems may have uncharacterized or hidden side channels due to manufacturing variations, environmental influences, or component aging. For example, even when optical isolators and wavelength filters are used to block Trojan Horse attacks, slight imperfections or unexpected leakage paths may still exist \cite{gisin2006trojan}.
    \item System Integration and Synchronization Issues:
    QKD protocols assume perfect synchronization between the transmitter and receiver, along with flawless operation of classical authentication channels. In reality, maintaining synchronization over long distances, calibrating equipment regularly, and ensuring the classical channel's security under realistic conditions all pose significant technical challenges that can reduce the overall effectiveness of the countermeasures.
\end{itemize}

These practical challenges illustrate why even carefully designed countermeasures can fail in real-world scenarios. They underscore the need for continuous improvement in device manufacturing, calibration techniques, and, perhaps most importantly, the development of a new generation of protocols that minimize the reliance on stringent assumptions about individual component behavior. DIQKD.

%%%%%%%%%%%%%%%%%%%%%%%%%%%%%%%%%%%%%%%%%%%%%%%%%%%%%%%%%

\section{Device Independent QKD (DIQKD)} \label{DIQKD}
Device-Independent Quantum Key Distribution (DIQKD) \cite{mayers1998quantum} is a promising approach to quantum cryptography that eliminates the need to trust the underlying infrastructure. DIQKD originated from Ekert's proposed E91 protocol \cite{ekert1991quantum}, in which an untrusted third party distributes an entangled pair of photons to the communicating parties, Alice and Bob. The communicating parties perform a Clauser-Horne-Shimony-Holt (CHSH) test on the entangled pair to certify the presence of a nonlocal correlation between them. These correlations ensure that no knowledge has been passed to an adversary, Eve.

\begin{figure*}
    \centering
    \includegraphics[width=0.75\textwidth]{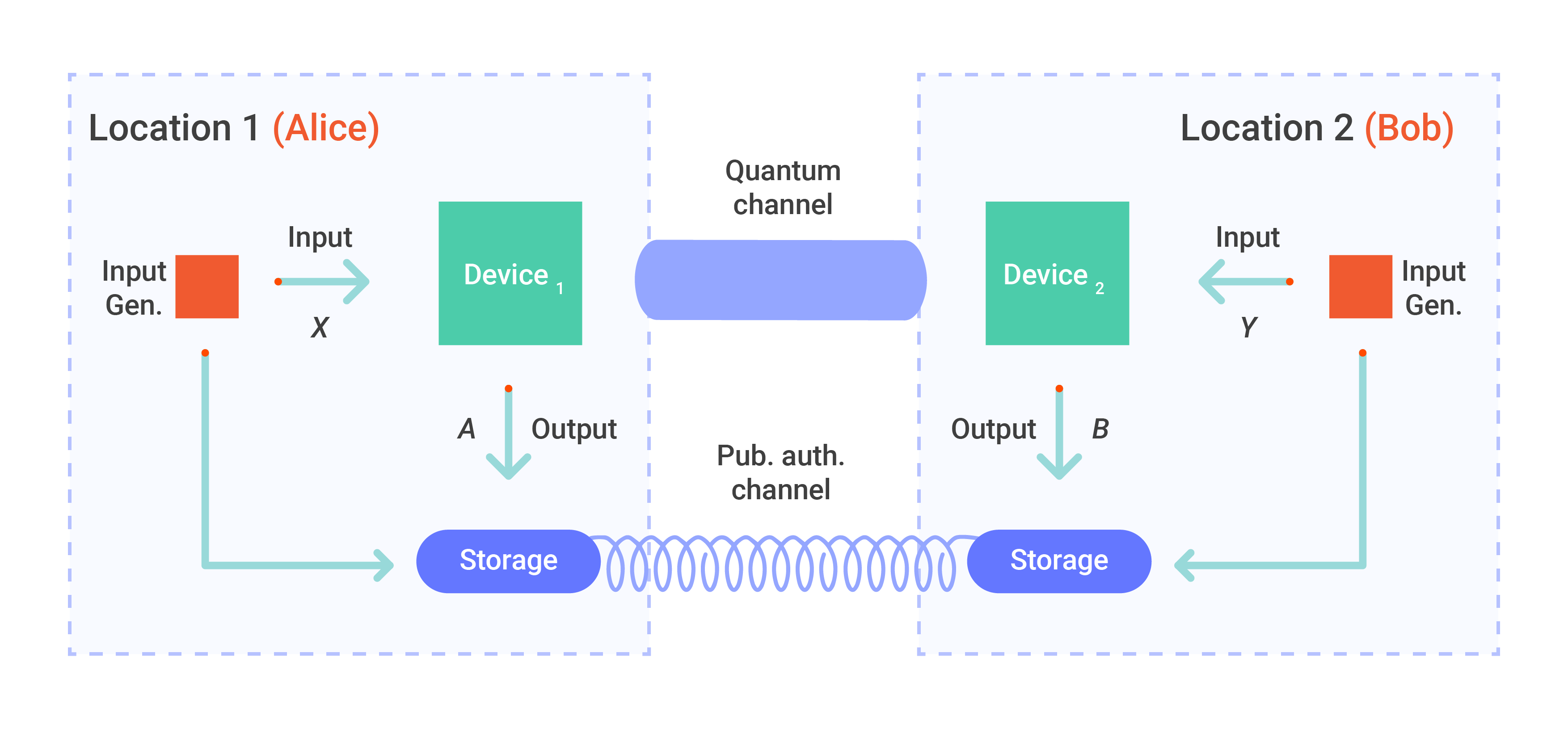}
    \caption{Schematic illustrating the basic setup of a DIQKD protocol. Alice and Bob are connected via two channels: 1) a quantum channel that enables the exchange of entangled quantum states or the transmission of quantum systems required for secure key generation, and 2) a publicly authenticated classical channel that facilitates the exchange of classical information necessary for key reconciliation and error correction. }
    \label{fig:enter-label}
\end{figure*}

\begin{figure}
    \centering
    \includegraphics[width=0.9\linewidth]{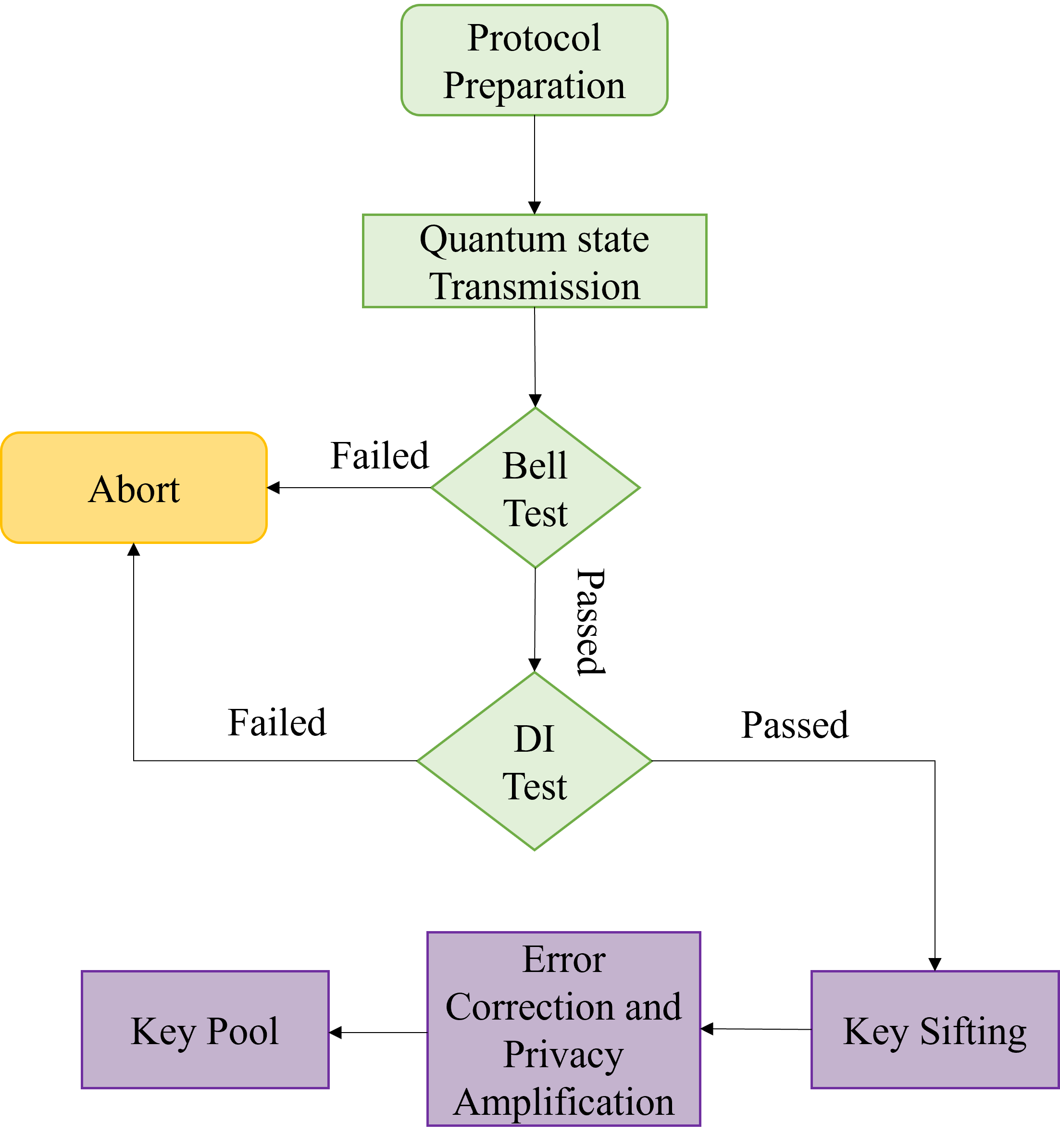}
    \caption{Overview of a DIQKD Protocol. The protocol begins with the preparation step, which includes the basis selection by Alice and Bob and the state preparation, after which the state is transmitted to Alice and Bob. Alice and Bob check if the Bell Test and Device-Independence check is passed; otherwise, the protocol is aborted. If the checks are passed, Alice and Bob perform the key sifting, error correction, privacy amplification, and key pooling steps to extract the final key.}
    \label{fig:diqkd_flowchart}
\end{figure}

Figure \ref{fig:widefig} provides an overview of key milestones in the evolution of DIQKD. The journey begins with Ekert's E91 protocol \cite{ekert1991quantum}, which offered the first insight into device independence by employing entangled photon pairs and Bell inequality tests to secure communication. A major breakthrough occurred in 2005 when Barrett \emph{et al.} \cite{barrett2005no} laid a foundational framework for DIQKD, although they did not explicitly coin the term ``Device-Independent". Their protocol relied on the no-signaling principle and the violation of Bell inequalities (using chained Bell inequalities) to guarantee security by bounding the information accessible to an eavesdropper, even under idealized conditions. Building on this framework, in 2006 Acin \emph{et al.} \cite{acin2006efficient,acin2007device} introduced a protocol based on the simpler and more efficient CHSH inequality, which further relaxed the requirement to trust the devices by deriving security solely from the observed nonlocal correlations. Together, these milestones highlight the shift from traditional, device-dependent security models to DIQKD, where security is certified through empirical statistical tests rather than assumptions about device integrity.

DIQKD addresses the limitations of traditional QKD by eliminating the need for trusted device models. Instead of relying on predefined device specifications, DIQKD uses Bell inequality tests to verify the entanglement between Alice and Bob. This approach is secure even in case the devices are faulty or untrusted: any attempt of eavesdropping, like injecting fictitious entangled states, memory attacks, or controlling the detectors, will disturb the quantum correlations and result in a failure to violate the Bell inequality. By certifying security through observed statistics rather than through assumptions about device behavior, DIQKD inherently resists all types of attack, including both main-channel and side-channel exploits, without requiring device-specific countermeasures. This paradigm shift represents a critical advancement toward quantum-safe security.

The CHSH inequality \cite{clauser1969proposed} is the generalized form of Bell's theorem, which proved to be decisive in experimental proofs of the local hidden variable given by Einstien, Rosen and Podolsky (known as EPR paradox) \cite{einstein1935can}. The CHSH inequality presented in \cite{clauser1969proposed} is:
\begin{align}
    S \equiv E(a,b)- E(a,b')+E(a',b)+E(a',b') \label{S}
\end{align}
The maximum bound which EPR predicted is \cite{einstein1935can}:
\begin{align}
    |S| \leq 2 \label{class_bound}
\end{align}
However, when using the entangled pairs for the protocol, the value of $S$ can reach up to $2\sqrt{2}$, thus violating the inequality in Eq. (\ref{class_bound}). This violation demonstrates the presence of some non-local correlations and will continue if the system is not disturbed by external sources. Otherwise, it will change to $\pm \sqrt{2}$, making the inequality un-violated. In the QKD system, we can exploit this fluctuation as a test for the presence of an adversary. Determining the value of $S$ necessitates collecting measurement outcomes from a designated setting. These measurement outcomes form a sequence of binary strings (i.e., the keys) generated by Alice and Bob. As the success rate is not 100\%, the keys are subject to an analysis known as quantum bit error rate (QBER). This is calculated by taking the probability of unsuccessful outcomes with the total outcomes.
\begin{align}
    QBER = Pr(a_0\neq b_2) \label{qber}
\end{align}
where $a_0$ and $b_2$ are the measurement outcomes of a specific measurement $A_0$ and $B_2$ performed by Alice and Bob on the entangled pair.  And by using the outcome in Eq. (\ref{qber}), the secret key rate is extracted as:
\begin{align}
    r \geq 1 - h(QBER) - \kappa (B_1 : E)
\end{align}
where $h(QBER)$ is the uncertainty in QBER, $\kappa (B_1 : E)$ is the Holevo bound on Bob's raw key and Eve. This bound measures how much information Eve has about Bob raw keys. To simulate Eve's maximum attack we need to maximize this Holevo bound. For a DIQKD system involving the CHSH game, the maximum Holevo quantity between Eve and Bob can be achieved by taking into consideration the value of $S$ from Eq. (\ref{S}) as:
\begin{align}
    \kappa (B_1 : E) \leq \Bigg( \frac{1 + \sqrt{(\frac{S}{2})^2-1}}{2} \Bigg) \label{holevo}
\end{align}
Equation (\ref{holevo}) shows that the Holevo information depends on the degree of CHSH violation, quantified by the value of $S$. Moreover, the secret key rate is determined by both, the Holevo information and the quantum bit error rate (QBER). This relationship allows the key rate to be plotted as a function of $S$ and QBER, providing valuable insight into how the degree of non-locality affects the overall security of the protocol. An overview of the DIQKD protocol is presented in Fig. (\ref{fig:diqkd_flowchart}).

Table \ref{diqkd_variants} summarizes the main variants of DIQKD along with their key properties, including assumptions made about devices, the level of trust required, the security guarantees provided, practical requirements, and the reliance on Bell tests. DIQKD implementations differ significantly in their underlying assumptions and security-certification techniques. In the following sections, we first examine \emph{fully device-independent quantum key distribution} (FDI-QKD), then discuss variants that relax its strict assumptions. Each approach presents unique advantages, limitations, and use cases. 
\begin{center}
\begin{table*} \label{diqkd_variants}
\caption{Variants of DIQKD along with their properties.}
 \begin{tabular}{| p{2.5cm} | p{2.5cm} |p{2.0cm}|p{3cm}|p{3cm}|p{2.5cm}|}

\hline
\textbf{DIQKD Type} & \textbf{Assumptions about Devices} & \textbf{Level of Trust} & \textbf{Security Guarantees} & \textbf{Practical Requirements} & \textbf{Reliance on Bell Tests} \\ 
\hline

Fully DIQKD \cite{vazirani2014fully} & No assumptions (both devices treated as black boxes) & No trust in any devices & Strongest security (based on Bell test violations) & Requires closing of all loopholes, high experimental rigor & Essential for security \\ 

\hline

One-Sided DIQKD \cite{branciard2012one} & Assumes one device (e.g., Bob's) is trusted and well-characterized & Trust in one device only & Strong security, but reliant on trust in one device & More practical than fully DIQKD, easier to implement & Important, but only for one device \\ 

\hline

Semi-DIQKD \cite{pawlowski2011semi} & Partial assumptions (e.g., dimensionality of systems) & Partial trust (e.g., limited to certain aspects like dimension) & Security depends on specific assumptions (e.g., dimensional witness) & Lower experimental demands, fewer loopholes to close & Can be minimized, depending on assumptions \\ 

\hline

Measurement-Device-Independent QKD \cite{lo2012measurement} & Assumes trusted preparation of quantum states & Trust in state preparation only & High security, resistant to measurement-device attacks & Feasible with current technology, does not require closing loopholes & Not required for security \\ 
\hline
\end{tabular}
\end{table*}

\end{center}
  \subsection{Fully Device-Independent QKD (FDI-QKD)}
  
When discussing device-independent QKD, we typically imply \emph{fully device-independent} (FDI-QKD). In particular, in FDI-QKD we do not need to trust the state-preparation or measurement devices beyond fundamental physical constraints (e.g., no faster-than-light signaling). This model makes assumptions about the fully trusted random number generator, the classically authenticated channel between the two parties, and the 100\%  secrecy of Alice and Bob's physical locations. FDI-DIQKD protocols are immune to a wide range of side-channel attacks that exploit imperfections or vulnerabilities in the devices due to these assumptions. However, these stringent requirements pose major challenges for the practical implementation of FDI-QKD. The most significant challenge is that fully DIQKD requires loophole-free Bell tests (discussed in Section \ref{chsh} and Table \ref{loophole}), which is difficult to achieve with today's technology. The first major proposal of FDI-QKD was made by Vazirani \emph{et al.} \cite{vazirani2014fully}, who provided a theoretical framework. However, their proposal relied on very strict idealized conditions, such as near-perfect Bell inequality violations, which are challenging to meet experimentally. Subsequently, Rodrigues and Lackey \cite{rodrigues_et_al:LIPIcs.TQC.2023.8} proposed a DIQKD protocol based on synchronous correlations and Bell's inequalities. Their approach aimed to mitigate the practical challenges by introducing a security assumption that closes the locality loophole: even an unbounded adversary, with only a slight uncertainty about the measurement bases, cannot produce nearly synchronous correlations that would mimic a maximal Bell violation. This refinement makes the protocol more robust under realistic experimental conditions while still maintaining the core advantage of DIQKD, namely, security derived solely from observed non-local correlations rather than from trust in the devices.

On the experimental side, Zahidy \emph{et al.} \cite{zahidy2024quantum} have recently demonstrated the use of a quantum dot-based single-photon source in a BB84 QKD protocol, paving the way for its application in FDI systems.

   \begin{figure}[H]
    \centering
    \includegraphics[width=1\linewidth]{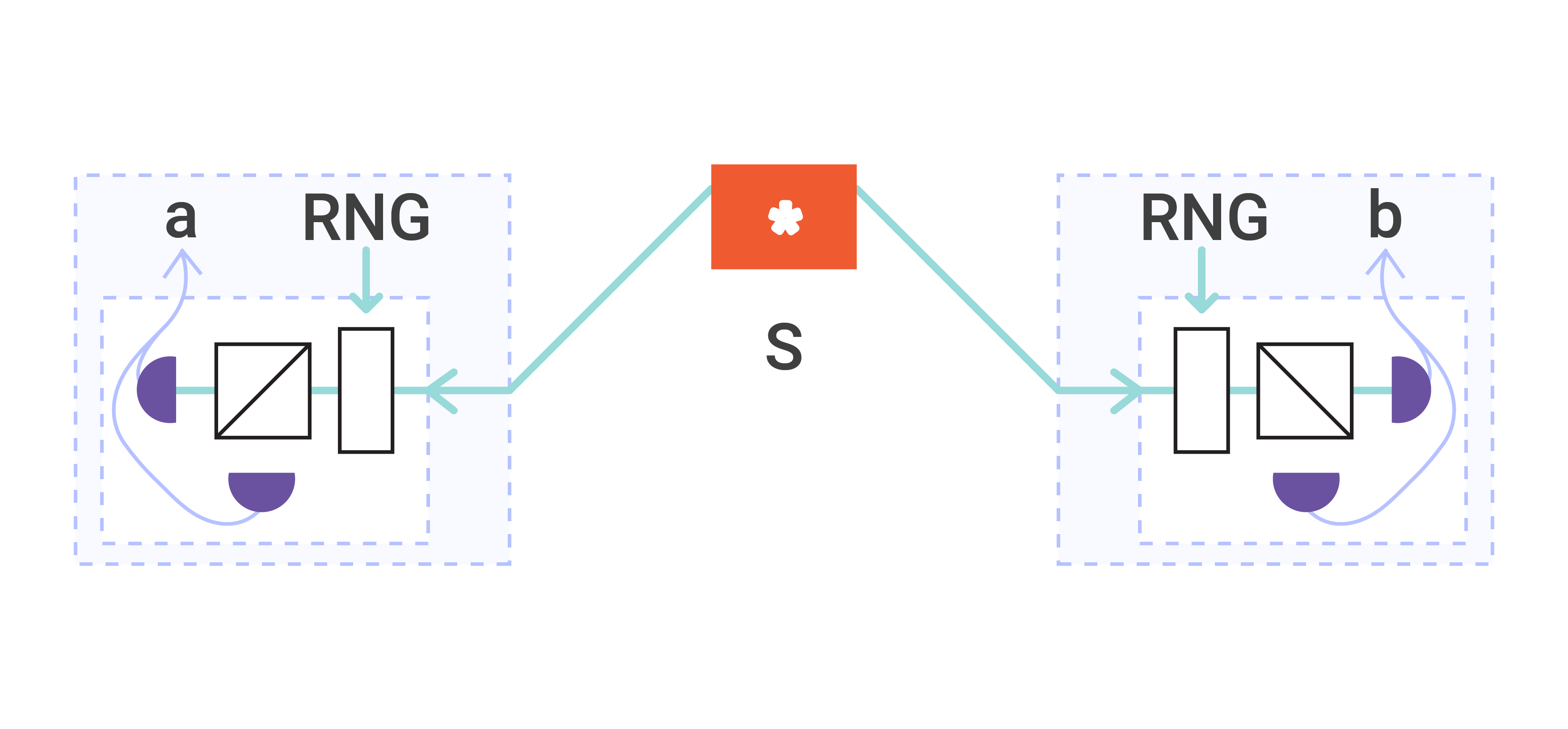}
    \caption{Visual representation of the Fully Device-Independent Quantum Key Distribution (FDI-QKD) protocol}
    \label{fig:enter-label1}
\end{figure}

\subsection{Partial and Alternative Approaches}
While FDI-QKD excludes the trust requirement in any quantum device, it faces serious practical challenges due to low efficiency, noise, and equipment quality constraints. To overcome these hurdles, less stringent protocols were proposed, including one-sided DI QKD, semi-DI QKD, and measurement-device-independent QKD (MDI-QKD). Each approach relaxes certain FDI-QKD assumptions to enhance feasibility. 
%The one-sided DI QKD trusts the single-party device, the semi-DI QKD assumes partially characterized devices, and MDI-QKD specifically removes trust in the measurement apparatus. The following subsections provide details about each protocol's security advantages and discuss their implementability in comparison to the fully device-independent paradigm.

\subsubsection{One Sided Device Independent QKD (1sDI-QKD)}
\begin{figure}[H]
    \centering
    \includegraphics[width=1\linewidth]{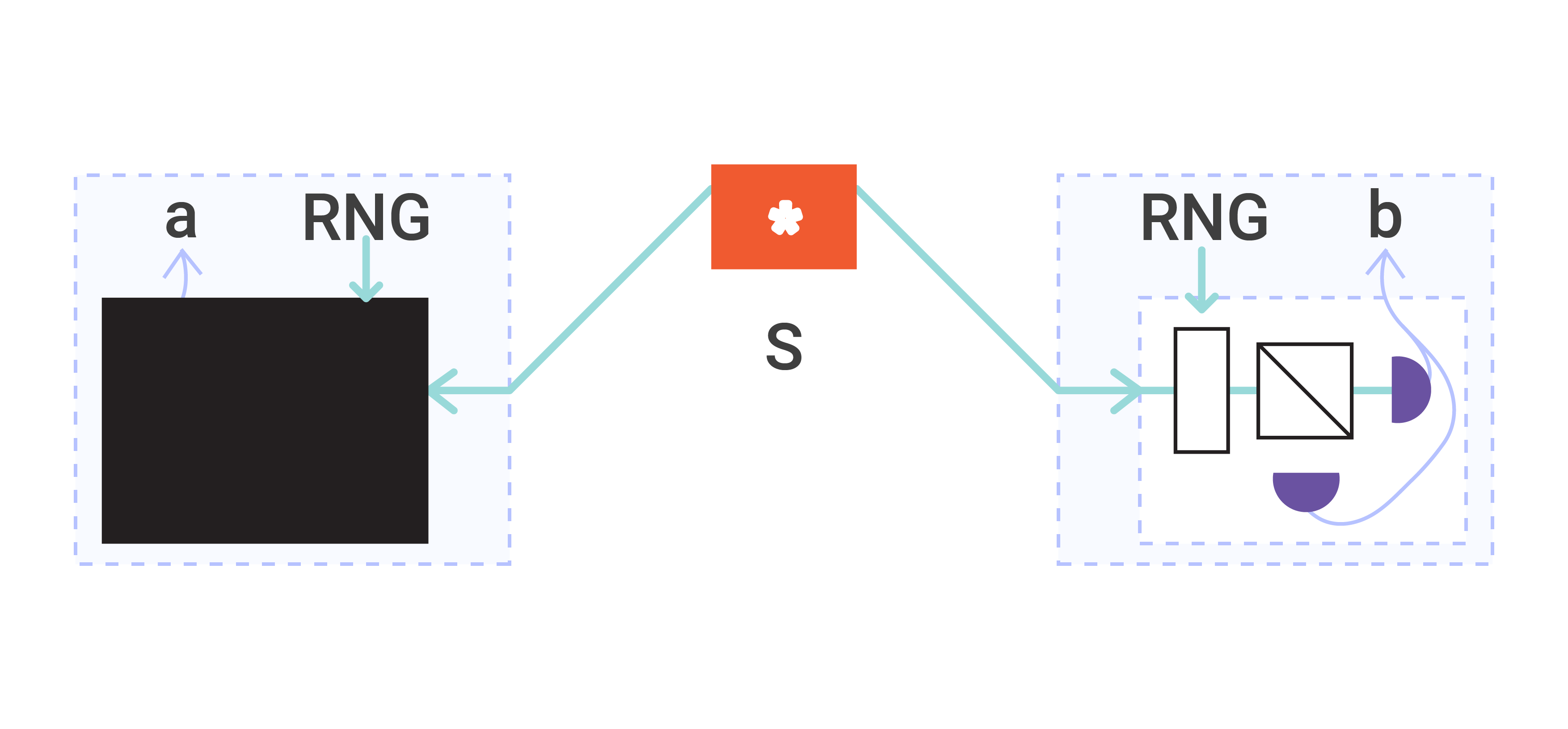}
    \caption{This schematic illustrates the framework of One-Sided Device-Independent Quantum Key Distribution (1sDI-QKD). In this protocol, the security of key distribution relies on the device-independence of one party's measurement apparatus (here Bob). In contrast, the other party's device, Alice, is assumed to be fully characterized or trusted.}
    \label{fig:enter-label1}
\end{figure}
Unlike FDIQKD, which requires Alice’s and Bob’s devices to be completely untrusted, 1sDI-QKD assumes that one of the communicating party devices is trustworthy (e.g., Bob's). This assumption simplifies the security analysis and reduces the experimental requirements, making the protocol more feasible in real-world scenarios. 1sDI-QKD was first introduced by Branciard \emph{et al.} \cite{branciard2012one} where they showed that in the case of on-sided DIQKD, the requirement for the generation of keys is much easier to meet than FDIQKD. Wang \emph{et al.} \cite{wang2013finite} analyzed the security of the one-sided protocol with finite resources by using the uncertainty relation for smooth entropies. Tomamichel \emph{et al.} \cite{tomamichel2013one} converted the conventional QKD BB84 protocol into 1sDI-QKD by making Alice's device trusted and Bob's untrusted. Walk \emph{et al.} \cite{walk2016experimental} provided an experimental implementation that uses Gaussian states to identify all protocols that can be 1sDI-QKD and their maximum loss tolerance. Xin \emph{et al.} \cite{xin2020one} showed that QKD between two independent parties could occur by resorting to 1sDI-QKD, which is also feasible in the presence of Gaussian states and measurements. Another approach to build a one-sided QKD is using EPR steering \cite{vallone2013einstein, pramanik2014fine,kogias2015quantification,chowdhury2015stronger, kaur2020fundamental, zhai2021manipulated,liu2022distillation,qars2024manipulating}.

\subsubsection{Semi-Device Independent QKD (SDI-QKD)}
 \begin{figure}[H]
    \centering
    \includegraphics[width=1\linewidth]{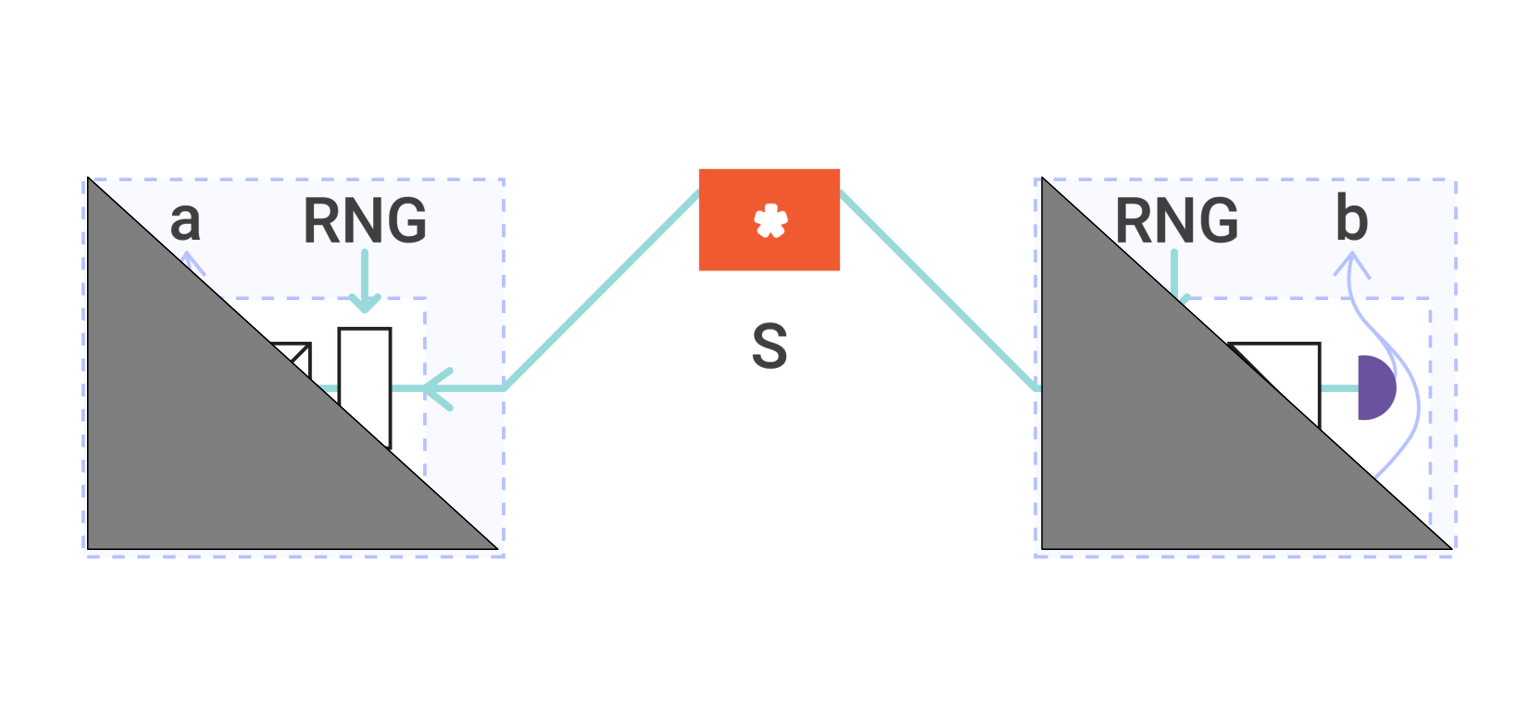}
    \caption{This schematic depicts the Semi-Device-Independent Quantum Key Distribution (SDI-QKD) protocol. In SDI-QKD, the security of the key distribution process is based on minimal assumptions about the devices, such as knowing the dimension of the quantum states (e.g., qubits). Unlike fully device-independent approaches, SDI-QKD does not require certification of the devices' trustworthiness but relies on operational constraints, such as the bounded dimension of the quantum systems.}
    \label{fig:enter-label1}
\end{figure}

In general, DIQKD exhibits two major challenges: (1) the difficulty of detecting loopholes and (2) its reliance on entanglement‐based protocols. Semidevice-independent QKD (SDIQKD) addresses these by reducing the security requirements to minimal, well-defined assumptions about devices, typically the dimension of the system or its ability to generate specific quantum states. One‐way prepare‐and‐measure protocols, such as BB84, involve Alice preparing a quantum state and sending it to Bob, who then performs a measurement. SDI‐QKD leverages these minimal assumptions to relax the full trust requirements of the devices, allowing one‐way prepare‐and‐measure schemes to be incorporated into a DIQKD framework. This combines the simplicity of prepare‐and‐measure with security based on limited, clear device assumptions \cite{pawlowski2011semi}.

Recent advances in semi-device-independent QKD (SDI-QKD) have addressed various aspects of reducing device trust while ensuring security. Yang \emph{et al.} \cite{wang2014security} laid a theoretical foundation by providing security proofs of SDI-QKD against collective attacks. Building on this, Woodhead \emph{et al.} \cite{woodhead2012semi,woodhead2016semi} implemented SDI-QKD protocols based on the BB84 scheme, demonstrating the practical viability of reducing device assumptions. Meanwhile, Chaturvedi \emph{et al.} \cite{chaturvedi2018security} investigated SDI-QKD protocols that utilize random access codes (RACs) as an alternative framework for key distribution, offering a distinct perspective on security. Zhou \emph{et al.} \cite{zhou2017finite} further contributed by proposing an SDI-QKD protocol with finite-resource security bounds, making the approach more applicable to real-world conditions. Finally, Jo \emph{et al.} \cite{jo2019semi} extended SDI-QKD into multi-party scenarios, highlighting its potential for broader network applications. Together, these studies illustrate a coherent progression—from robust security proofs and experimental demonstrations to practical security bounds and multi-user extensions—underscoring the evolution of SDI-QKD as a promising approach to quantum key distribution.

\subsubsection{Measurement Device Independent QKD (MDI-QKD)}
\begin{figure}[H]
    \centering
    \includegraphics[width=\linewidth]{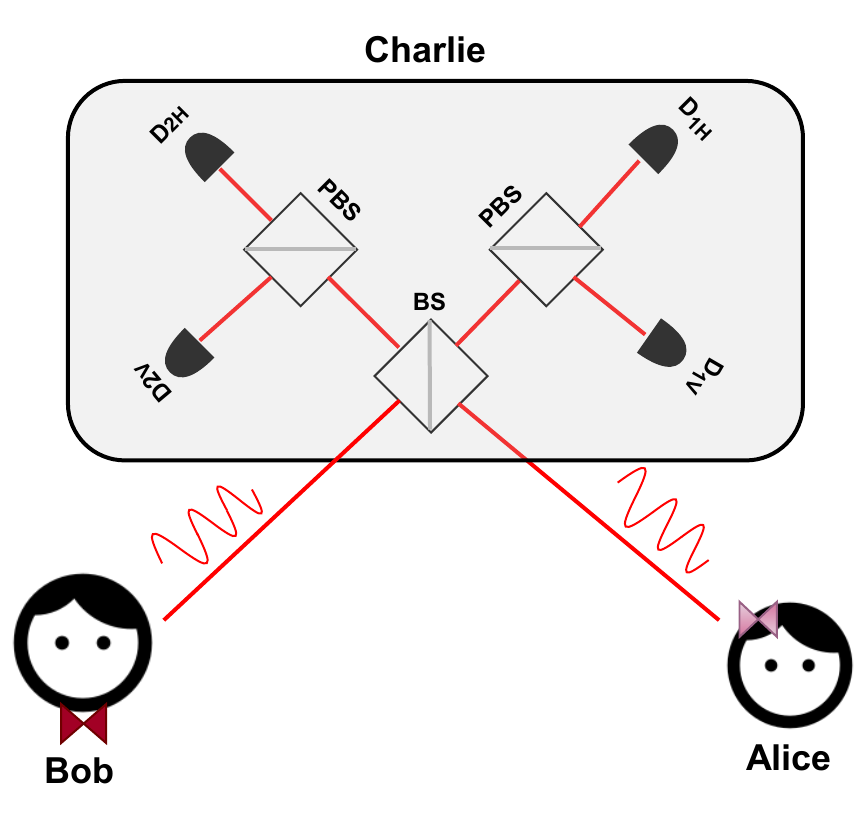}
    \caption{This schematic depicts the Measurement-Device-Independent Quantum Key Distribution (MDI-QKD) protocol, an advanced QKD protocol designed to eliminate vulnerabilities associated with the detection process. In MDI-QKD, Alice and Bob do not directly measure their exchanged quantum states. Instead, they send their prepared quantum states to an untrusted third party, Charlie, who performs a joint measurement and announces the results over a public classical channel.}
    \label{fig:enter-label2}
\end{figure}
In contrast to 1sDI-QKD and SDI-QKD, which relax specific assumptions of FDI-QKD, MDI-QKD adopts a fundamentally different approach. MDI-QKD eliminates the need to trust the measurement devices by utilizing an untrusted intermediary for all measurements. The measurement process is outsourced to a third party ``Charlie", thus eliminating the threat posed by the malicious measurement devices.
MDI-QKD was discovered by Lo \emph{et al.} \cite{lo2012measurement} while they trying to solve the detector side-channel attacks. 

With the currently available technology, MDI-QKD is the most practical and feasible solution as it offers high key rates at long distances compared to any other type of DIQKD \cite{xu2014measurement}. The first experimental realization of MDI-QKD was performed by Tang \emph{et al.} \cite{tang2016experimental}, where they were able to transfer secure keys over 40km with imperfect sources, which led to many subsequent experiments \cite{wang2016experimental,yin2016measurement,tian2022experimental,zhou2020experimental,valivarthi2019measurement,roberts2017experimental,hu2018measurement}. 

MDI-QKD protocol has also been realized for high-dimensional measurements. For example, Cui \emph{et al.} \cite{cui2019measurement} proposed a MDI-QKD setup that measures qudits that were encoded in high dimensions. Dellantonio \emph{et al.} \cite{dellantonio2018high}, shows that high-dimensional qudits encoding improves the performance of many current MDI-QKD implementations. This proved unconditional security even for weak coherent pulses with decoy states. Li \emph{et al.} \cite{li2018high} proposed a high-dimensional encoding scheme in MDI-QKD by a spatial-temporal mode conversion circuit and quantum state fusion operation,  simulation of this proposal shows that by this new scheme, the key generation rate increases by 2 to 10 times and transmission distance increases by 6 to 60 km for different polarization misalignment errors. Yang \emph{et al.} \cite{yang2021feasible} proposed a high-dimensional MDI-QKD protocol in multiple degrees of freedom DOFs involving a linear-optical hyperentangled Bell state analysis (HBSA). The linear-optical HBSA made it feasible under current experimental technology. The key generation rate of this protocol was eight times better than the key rate of the original MDI-QKD protocol. Sekga \emph{et al.} \cite{sekga2023high} proposed a high-dimensional MDI-QKD, which employed biphotons to encode information. Cao \emph{et al.} \cite{cao2020long}, developed a robust adaptive optics system with high precision time synchronization and frequency locking between independent photon sources located far apart, which realized the first free-space MDI-QKD over a 19.2 km urban atmospheric channel (paving the way for the satellite based MDI-QKD). Yan \emph{et al.} \cite{yan2021measurement}, provided an efficient approach to increase the key generation rate of MDI-QKD by adopting multiple degrees of freedom (DOFs) of single photons to generate keys. In the engineering side, Wei \emph{et al.} \cite{wei2020high} developed a 1.25 GHz silicon chip-based MDI-QKD system that enhanced the finite secret key rate to 31 bit/s and due to its miniaturizing, low cost, and compatibility with CMOS microelectronic, it can serve in secure quantum networks in the future. Jiang \emph{et al.} \cite{jiang2021higher} proposed MDI-QKD using a double scanning method to improve the key rate. Lai \emph{et al.} \cite{lai2018high} proposed MDI-QKD protocol with Fibonacci-valued and Lucas-valued orbital angular momentum (OAM) entangled states in free space to enhance key rate and coding capacity.

\subsubsection{Detector-Device-Independent QKD (DDI-QKD)}
MDI-QKD setups traditionally suffer from low key rates, and to overcome this challenge, Detector-Device-Independent QKD (DDI-QKD) was proposed \cite{ma2011improved}.  DDI-QKD protocols employ two-qubit single-photon Bell state measurement (BSM) rather than the two-photon BSM used in MDI-QKD setups.  This modification simplifies the experimental setup and reduces the need for extensive post-processing of large data blocks required for finite-key security against general attacks by streamlining the measurement process.  However, DDI-QKD setups are vulnerable to malicious attacks.  For example, the combination of blinding attacks with intrinsic imperfections of single-photon detectors proposed by Wei \emph{et al.} \cite{wei2017feasible}, shows that Eve can obtain the secure key without being detected by this attack.  Furthermore, Sajeed \emph{et al.} \cite{sajeed2016insecurity} demonstrated that DDI-QKD is indeed vulnerable against side-channel attacks.  No significant advancements have been reported on this MDI-QKD variant.

In exploring the diverse landscape of DIQKD, we have examined various approaches, each with its unique set of assumptions, trust levels, and security guarantees. These methods, whether fully device-independent, one-sided, or semi-device-independent, illustrate the flexibility and adaptability of quantum cryptography in addressing real-world challenges. As we shift our focus to the underlying mechanisms that enable these protocols, it becomes evident that non-local games play a crucial role in establishing the security foundations of DIQKD. These games, rooted in the principles of quantum mechanics, serve as the bedrock for verifying the quantum correlations necessary for secure key distribution. In the following section, we look through the significance of these games, exploring how they are used to ensure the robustness of DIQKD protocols and their critical role in testing the limits of quantum theory.

%%%%%%%%%%%%%%%%%%%%%%%%%%%%%%%%%%%%%%%%%%%%%%%%%%%%%%%%%

\section{Quantum Games} \label{games}

Besides their entertainment value, games are also perceived as an intuitive way to find solutions to complex problems. The field that enables us to perceive it in such a way is called ``Game Theory". In this field, games are considered a powerful tool for solving complex problems in various domains, including economics, social sciences, and computer science. In quantum mechanics, quantum games extend classical game theory to quantum scenarios, where quantum properties (such as entanglement) provide new dimensions for problem solving.
Quantum games are commonly used to analyze secure communication protocols against eavesdroppers, Eve, to calculate the probability that Eve will be able to predict Alice's measurement.

\subsection{Two-Party Game}\label{game1}
We consider the case where Eve prepares qubits and sends them to Alice. Eve can make measurements on the qubits during the transmission phase. This scenario can be turned into a simple guessing game as follows:
\begin{enumerate}
    \item Eve prepares a qubit $\rho_{A}$ and sends it to Alice.
    \item Alice chooses a random bit, $\Theta\in [0,1]$, where $\Theta$ represents the choice of measurement basis.
    \item If $\Theta = 0$, Alice measures her qubit $\rho_{A}$ in the standard basis. If $\Theta = 1$, Alice measures in the Hadamard basis.
    \item Alice records the outcome of the measurements $X \in [0,1]$.
    \item Alice publicly announces $\Theta$.
    \item If Eve correctly guesses the outcome of the measurement $X$, she wins the game.  
\end{enumerate} 
This game illustrates that, in this scenario, Eve prepares the qubits while Alice generates a key. In conventional QKD settings, the party involved in key preparation is typically trusted. However, in the game, Eve cannot predict Alice’s measurement outcomes with 100\% accuracy—a limitation that arises from the uncertainty principle. This is quantified by a bound on the average probability that Eve correctly guesses $X$:
\begin{align}  
    P_{guess}(X|\Theta) = p(\Theta=0).P_{guess}\label{guess1}(X|\Theta=0)\\+p(\Theta=1).P_{guess}(X|\Theta=1) \notag
\end{align}
As we can see in the game, Alice has only two outcomes for $\Theta$; this makes it a uniform probability, and hence $p(\Theta=0) = p(\Theta=1)$ $=1/2$. Hence:
\begin{align}
    P_{guess}(X|\Theta) = \frac{1}{2}\big[P_{guess}\label{guess2}(X|\Theta=0)\\+P_{guess}(X|\Theta=1)\big] \leq c\notag
\end{align}
As Eve holds no additional information regarding $X$, except Alice's basis choices, the value of $c$ will always be less than 1.

It is clear that Eve cannot guess the key with 100\% certainty. However, the risk of compromised communication remains. Hence, Eve may attempt to maximize her probability of correctly guessing Alice's outcomes $X$. 
She can do this by picking the largest eigenvalue of the combined eigenvector of the standard and Hadamard basis $\big(\ket{0}\bra{0}+\ket{+}\bra{+}\big)$. The largest eigenvalue is $\lambda_{max}=1+\frac{1}{\sqrt{2}}$, substituting it in Eq. (\ref{guess2}) yields $P_{guess}(X|\Theta)=\frac{1}{2}+\frac{1}{2\sqrt{2}}$. This means that the winning probability of Eve in this game is 0.85 or 85\% with the min-entropy \cite{konig2009operational} \footnote{The relation between guess probability and min-entropy is $H_{min}(X|\Theta)=-log P_{guess}(X|\Theta)$} $H_{min}(X|\Theta)$ = 0.22. This value of min-entropy does not guarantee a safe environment for secure communication, so we can say that the uncertainty principle helps us by making it difficult for Eve to guess the measurement outcomes, but is insufficient to tackle the eavesdropper problem.

To facilitate a secure environment, we must limit Eve's knowledge of Alice's outcome $X$. For this, we use another property of quantum mechanics known as \textit{Entanglement}. Entanglement is monogamous \cite{yang2006simple} in nature, if we find a way for two or more communicating parties to be largely entangled, we can reduce Eve's knowledge and consequently increase the min-entropy value. In the current guessing game scenario, unfortunately, we cannot incorporate both properties \textit{i.e.} entanglement and uncertainty principle, as this is a two-party scenario where one party is the adversary (establishing entanglement between Alice and Eve means that Eve will always guess all of Alice's outcomes correctly).

\subsection{Three-Party Game}\label{game2}
Now we involve a third party named Bob, whom Alice trusts. In this game, Alice and Bob are entangled, which increases the min-entropy relative to Eve’s potential information. To do so, Alice and Bob need to perform entanglement tests (see Section \ref{Nonlocal}). If the results of these tests show a high level of entanglement, then monogamy of entanglement will prevail between Alice and Bob. This game is summarized as follows:
\begin{enumerate}
    \item Eve prepares a three-party state $\rho_{ABE}$ and sends qubits $A$ and $B$ to Alice and Bob, respectively.
    \item Alice chooses a random bit $\Theta$ where $\Theta\in[0,1]$
    \item If $\Theta=0$, Alice measures $\rho_A$ in the standard basis. Otherwise, she measures in the Hadamard basis.
    \item She obtains the measurement outcome $X \in [0,1]$ and records it.
    \item Alice publicly announces her basis choices $\Theta$.
    \item Eve and Bob both measure their qubits ($\rho_B $ and $\rho_E$) with the basis Alice announced $\Theta$
    \item  Bob and Eve win the game if they guess $X$ correctly $X_E=X=X_B$.
\end{enumerate}
The winning probability of Bob and Eve is given by:
\begin{align}
    p_{BE} = \sum_{\Theta\in[0,1]} \notag p_{\Theta}\text{Tr}\big[\rho_{ABE}\Big(\sum_{x\in[0,1]}\ket{x}\bra{x}_{\Theta}^{A}\\\otimes )\ket{x}\bra{x}_{\Theta}^{B}\otimes M_{x|\Theta}^{E}\Big)\big] \label{tri1}
\end{align}

Here, superscripts $A$, $B$, and $E$ represent the measurements performed by Alice, Bob, and Eve systems, respectively. Again, the choice of the basis is two, hence the probability of $p_{\Theta}=\frac{1}{2}$. This modifies Eq. (\ref{tri1}) as:
\begin{align}
    p_{BE} = \frac{1}{2}\sum_{\Theta\in[0,1]} \notag \text{Tr}\big[\rho_{ABE}\Big(\sum_{x\in[0,1]}\ket{x}\bra{x}_{\Theta}^{A}\\\otimes )\ket{x}\bra{x}_{\Theta}^{B}\otimes M_{x|\Theta}^{E}\Big)\big] \label{tri2}
\end{align}

The strategy of Eve's measurement $M_{x|\Theta}^{E}$ is unknown to us, along with the quantum state $\rho_{ABE}$ which Eve has prepared. Considering the worst-case scenario, in which Eve has unlimited resources, we can formulate the winning probability. Using relevant algebra tools \cite{tomamichel2013monogamy}, we can calculate the winning probability of this game, which will be 0.85 or 85\%. In Section \ref{game1}, we also obtain the same winning probability for Eve. Now the question arises of how this version of the game is better than the game presented in Section \ref{game1}. We refer to the discussion in the last part of the former section to answer this. It is discussed that if Eve tries to entangle with Alice, she could know the outcomes with 100\% certainty. However, the game presented in this section limits Eve's power to gain information about Alice's outcomes even when she can entangle herself with Alice. Including the third party weakens the entanglement power between Alice and Eve, hence increasing the min-entropy value.

\subsection{Non-local Games}\label{Nonlocal}
Non-local games are strategic scenarios in which spatially separated players (who cannot communicate during the game) aim to maximize a shared winning condition. They typically exploit the phenomenon of quantum entanglement, allowing players to achieve correlations that exceed any classical limits. We observe that for many guessing games, quantum strategies yield a higher winning probability. We can also relate this winning probability to the presence of entanglement, thus making it the ultimate test for checking the entanglement between the communicating parties. While several non-local games have been proposed as promising alternatives to the traditional CHSH game in DIQKD systems (such as the Mermin-Peres Magic Square, GHZ, Monty Hall, and RGB games), it is important to note that, except for CHSH, these games are very difficult to implement in practical scenarios because they operate under ideal conditions. Following this, we will now discuss two major games that are used in DIQKD.

\subsubsection{CHSH game} \label{chsh}
CHSH game is based on the famous CHSH inequality. In this game, two bits, $x$ and $y$, are sent to Alice and Bob, respectively. The values of $x$ and $y$ are chosen uniformly at random, where $p(x=0)=p(x=1)=p(y=0)=p(y=1)=\frac{1}{2}$. Before the game starts, Alice and Bob are allowed to agree on a strategy, but no communication is allowed during the game. Next, Alice and Bob send their outcomes $a$ and $b$, respectively, to the referee, who checks whether the winning condition is met. Alice and Bob win the game if the following condition is satisfied:
\begin{align}
    x\cdot y&=a+b\ (\text{mod}\ 2) \label{CHSH}
\end{align}
where $\cdot$ and $+$ represent the AND and OR operators. We are only interested in the wining of Alice and Bob therefore, we define the winning probability:
\begin{align}
    P_{CHSH}^{win}&= \sum_{x,y}P(x,y)\sum_{a,b}P(a,b|x,y)\delta(x\cdot y,a+b\ (\text{mod}\ 2))
\end{align}
Since both $x$ and $y$ are chosen uniformly at random from $\{0,1\}$, there are four equally likely outcomes $(00, 01, 10, 11)$. Consequently, the probability of any particular pair $(x,y)$ is $P(x,y)= \frac{1}{4}$.
% The expression $P(x,y)= \frac{1}{4}$, because out of four possible outcomes of $x$ and $y$ (00, 01, 10, 11) out of which Alice and Bob are given one pair randomly. 

The objective of the game is for Alice and Bob to agree on a strategy that will maximize their probability of winning the game. % of how Alice and Bob ensure that they win the game every time. In principle, they have two ways to deal with this situation $(i)$ they can apply a classical strategy, or $(ii)$ they can apply quantum strategy. 

%\vspace{0.2cm}

\paragraph{Classical winning strategy.} In the classical approach to the CHSH game, Alice and Bob prearrange a deterministic strategy based solely on their inputs \(x\) and \(y\) to maximize their chances of satisfying the winning condition.  Let $a$ and $b$ be functions of $x$ and $y$, respectively; if Alice and Bob are given a bit in which $x$ or $y$ = 0, then according to the AND operation in the condition $x\cdot y=0$ (includes the bits 00, 01, 10). To achieve $a + b (\text{mod} 2) = 0$, there is a chance of 3 out of 4 times that  $x\cdot y=0$, if they receive $x=0$ or $y=0$. The strategy of returning bits can be predetermined as $a=b = 0$, which will satisfy the condition given in Eq. (\ref{CHSH}) and hence win the game. However, there is a chance that when they receive the bit $x=y = 1$, they will lose the game as they have no information about the bits of each other and will end up not satisfying the inequality. Thus, the maximum winning probability is:
\begin{equation}
    P_{CHSH}^{(C),win} = \frac{3}{4} = 75\%
\end{equation}
where $C$ represents the classical strategy. %Now, we will check that by applying the quantum strategy, we can improve our winning condition.

\paragraph{Quantum winning strategy.} In this strategy, Alice and Bob share an entangled pair before the game starts. 
\begin{align}
    \ket{\Psi}_{AB} = \frac{1}{\sqrt{2}}\big(\ket{0}_A\ket{0}_B + \ket{1}_{A}\ket{1}_B\big)
\end{align}

The subscripts $A$ and $B$ represent the qubits that Alice and Bob possess, respectively. According to the bits $x$ and $y$ provided by the third party (i.e., Charlie), they choose the bases accordingly. For Alice, it is straightforward, if $x=0$, she uses the standard basis $(\ket{0},\ket{1})$ and if $x=1$, she uses the Hadamard basis $(\ket{+},\ket{-})$. 
On the other hand, the bases Bob chooses are more interesting. If $y=0$, he measures in:
\begin{align*}
    \ket{b_1} &= \cos(\frac{\pi}{8})\ket{0}+\sin(\frac{\pi}{8})\ket{1},\\
     \ket{b_2} &= -\sin(\frac{\pi}{8})\ket{0}+\cos(\frac{\pi}{8})\ket{1}
\end{align*}
and when $y=1$, Bob measures in:
\begin{align*}
     \ket{c_1} &= \cos(\frac{\pi}{8})\ket{0}-\sin(\frac{\pi}{8})\ket{1},\\
     \ket{c_2} &= \sin(\frac{\pi}{8})\ket{0}+\cos(\frac{\pi}{8})\ket{1}
\end{align*}
To determine the winning probability, we consider all possible combinations of $x$ and $y$. 

\begin{itemize}
    \item when $x=y=0$:
\begin{align*}
    p_{win|00} &= p(a=0,b=0|x=0,y=0) \\
    &+ p(a=1,b=1|x=0,y=0)\\
    &= |\braket{0_Ab_{1B}}{\Psi_{AB}}|^2 + |\braket{1_Ab_{1B}}{\Psi_{AB}}|^2 \\&+|\braket{0_Ab_{2B}}{\Psi_{AB}}|^2+|\braket{1_Ab_{2B}}{\Psi_{AB}}|^2\\
    &=  2\Big|\frac{1}{\sqrt{2}}\cos^2(\frac{\pi}{8})\Big|^2= \cos^2(\frac{\pi}{8})
\end{align*}
    \item when $x=0,y=1$, $x=1,y=0$, and $x=1,y=1$:
\begin{align*}
    p_{win|01} = p_{win|10} = p_{win|11} &= \frac{1}{4} \sum_{x,y} p_{win|xy}\\
    &=\cos^2(\frac{\pi}{8}) \approx \text{0.85}
\end{align*}
\end{itemize}
where 1/4 is the probability $p(x,y)$.

Most of the DIQKD protocols today are based on CHSH games. However, there is a growing trend among researchers to use various non-local games to obtain a better winning probability. An example of such an alternative game is the Mermin-Peres Magic square game.

\subsubsection{Mermin-Peres Magic Square game}
The Mermin-Peres Magic Square game \cite{peres1990incompatible,mermin1993hidden} is an updated version of the Magic Square game that is both simple and can efficiently establish non-locality between two parties. In this game, two players, Alice and Bob, are given a $3\cross3$ square grid. Their task is to provide an input in $\{+1,-1\}$. Alice provides her results in a row, and Bob provides his results in a column. To win the game, they must satisfy two rules: 
\begin{enumerate}
    \item The product of the row must be equal to +1, and the product of the column must be equal to -1
    \item The cell that coincides with the row and the column should contain the matching numbers
\end{enumerate}
Before starting the game, Alice and Bob are given unlimited time to determine their strategy to win the game, but as soon as the game starts, they are not allowed to communicate. At the start of the game, the referee randomly selects the row number $x$ and the column number $y$ for Alice and Bob to fill. Alice and Bob respond with the results $a_{c}^{x}$ and $b_{c}^{y}$ separately, where $c$ is the cell number with $c\in \{0,1,2\}$. After providing their respective outcomes, the referee analyses the game's results based on the game rules and determines the winning probability. %There are two strategies to deal with this game, the first one being a classical strategy and the second one being a quantum strategy. Let us first go through the classical strategy.

\paragraph{Classical winning strategy.} Alice and Bob are given ample time before the start of the game to discuss how they can fill in their respective rows and columns to meet the requirement for winning the game. However, there are numerous ways to strategize their win in this game, but none guarantee a win with 100\% probability. The strategy that gives the maximum winning probability is as follows: before starting the game, Alice decides to put \textbf{1} in the first two rows and in the last row, she outputs \textbf{-1} in the first two cell and \textbf{1} in the last cell. In this way, she satisfies the first of the two conditions. Similarly, Bob fills all three columns with $\textbf{\{1, 1, -1\}}$, which also satisfies his first rule. The strategies for Alice and Bob are shown in Table \ref{CstratMPG}.
\begin{table}[H]
\centering
\begin{logicpuzzle}[rows=3,columns=3,width=3cm,scale=0.75,fontsize=Large]
\valueV{\color{ForestGreen}{2},\color{ForestGreen}{1},\color{ForestGreen}{0}}
\sumH{{},{},{}}
\valueH{{},a) Alice,{}}
    \setrow{3}{1,1,1} 
\setrow{2}{1,1 ,1 }
\setrow{1}{-1,-1,1} 
\framepuzzle[black!50] \hspace{.5cm}
\end{logicpuzzle}%
\hspace{1.5cm}
\centering
\begin{logicpuzzle}[rows=3,columns=3,width=3cm,scale=0.75,fontsize=Large]
\sumH{\color{blue}{0},\color{blue}{1},\color{blue}{2}}
\valueH{{},b) Bob,{}}
    \setrow{3}{1,1,1}
\setrow{2}{1,1 ,1 }
\setrow{1}{-1,-1,-1}
\framepuzzle[black!50]
\end{logicpuzzle}
\caption{a) Alice's pre-determined classical strategy. b) Bob's pre-determined classical strategy \label{CstratMPG}}
\end{table}

\begin{table}[H]
\centering
\begin{logicpuzzle}[rows=3,columns=3,width=4cm,scale=1,fontsize=Large]
\sumH{\color{blue}{0},\color{blue}{1},\color{blue}{2}}
\valueV{\color{ForestGreen}{2},\color{ForestGreen}{1},\color{ForestGreen}{0}}
% \valueH{{},b) Bob,{}}
    \setrow{3}{{1,1},{1,1},{1,1}}
\setrow{2}{{1,1},{1,1} ,{1,1} }
\setrow{1}{{-1,-1},{-1,-1},\color{red}{{1,-1}}}
\framepuzzle[black!50]
\end{logicpuzzle}
\vspace{0.5cm}
\caption{Combined result of classical strategy for MPG game} \label{combMPG}
\end{table}
A review of the combined outcomes in Table \ref{combMPG} reveals that no matter which values the input they receive, they will win the game except for one input $(x = 2, y = 2)$. Therefore, we conclude that the maximum winning probability for this game using a predetermined classical strategy is $\omega_{max}=\frac{8}{9} \approx \text{0.89}$. Depending upon the number of rounds $n$, the winning probability will become:
\begin{align}
    \omega_{max}^{n}=\Big(\frac{8}{9}\Big)^{n}
\end{align}

\paragraph{Quantum winning strategy.} The quantum strategy includes using two Bell pairs, which Alice and Bob share.
\begin{align}
    \ket{\Psi}_{AB} = \frac{\big(\ket{0}_{A1}\ket{0}_{B1} + \ket{1}_{A1}\ket{1}_{B1}\big)}{\sqrt{2}}\otimes \notag \\\frac{\big(\ket{0}_{A2}\ket{0}_{B2} + \ket{1}_{A2}\ket{1}_{B2}\big)}{\sqrt{2}} \label{max-bell}
\end{align}
The outcomes of Alice and Bob are obtained through the measurement of these Bell pairs. The correlation between these Bell pairs provides supremacy against any kind of classical strategy.

The optimal strategy to win this game is to exploit a certain order of measurement on the maximally entangled state presented in Eq. (\ref{max-bell}). The measurements are set in a $3\cross3$ table presented below:
\begin{table}[H]
\centering
\begin{logicpuzzle}[rows=3,columns=3,width=6cm,scale=2,fontsize=Large]
    \setrow{3}{$I\otimes Z$,$Z\otimes I$,$Z\otimes Z $}
\setrow{2}{$X\otimes I$,$I\otimes X$ ,$X\otimes X$ }
\setrow{1}{$-X\otimes Z$,$-Z\otimes X$,$Y\otimes Y$}
\framepuzzle[black!50]
\end{logicpuzzle}
\vspace{0.5cm}
\caption{Measurement settings of the optimal quantum strategy for MPG game. X, Y, Z are Pauli spin operators.} \label{QMPG}
\end{table}
In Table \ref{QMPG}, each row and each column commute, which means that the order of the measurement does not matter. For all rows $x$, the product of Alice's output will always be +1 as $\prod_{k}{M_{A}^{(x,k)}}= I$ and similarly for columns $y$ the product will be -1 as $\prod_{k}{M_{B}^{(y,k)}}= -I$. This satisfies our first condition for the MPG game. For the second condition, the operators in each cell of the rows and columns will give us eigenvalues +1 or -1 with $\ket{\Psi}_{AB}$. Those eigenvalues are considered as the outputs of Alice and Bob. To verify that Alice and Bob are always winning in each cell, we calculate the probability $\bra{\Psi_{AB}}M_{A}^{(x,y)} \otimes M_{B}^{(y,x)} \ket{\Psi_{AB}}$. More detailed calculations are presented in the supplemental material of Xu \emph{et al.} \cite{xu2022experimental}. This probability always gives the value 1, which confirms that the results in that specific cell match. This makes the winning probability of the quantum strategy 100\%, due to the pseudotelepathic \cite{gisin2007pseudo} power of the EPR pair.

 \subsubsection{Other non-local games}
 In general, for a game to be useful for key distribution, it must 1) be played between multiple players, 2) exhibit a quantum advantage in the winning strategy, and 3) be fair and symmetric. With this in mind, other potential games (apart from CHSH and MPG) may have the potential to be used in DIQKD.
 \begin{enumerate}
    \item \textit{GHZ game:} In the GHZ game, three parties, Alice, Bob, and Charlie, independently select the input bits \(x,y,z\in\{0,1\}\) under the constraint \(x\oplus y\oplus z=0\) and return an output bit \(a,b,c\), and they win if 
\[
a \oplus b \oplus c = x \lor y \lor z.
\]
Classically, the optimal success probability is \(\tfrac{3}{4}\), while sharing the GHZ state \(\tfrac{1}{\sqrt{2}}(\ket{000} + \ket{111})\) and performing appropriate local measurements yields unit success probability. QKD protocols using this game appear in \cite{toyota2010key,zhang2023rational}, and its DIQKD application was first proposed in \cite{basak2019device}.

     \item \textit{Monty Hall game:} In the quantum Monty Hall game \cite{flitney2002quantum}, a host (Monty) and a contestant (Alice) interact through a three-level quantum system (qutrit). Monty conceals a prize behind one of three doors (encoded as a qutrit state), and Alice initially selects a door by preparing the corresponding qutrit. Monty then 'opens' one of the two unchosen doors, performing a projective measurement to reveal no prize, and offers Alice the option to stay with her original choice or switch to the remaining closed door. Classically, the optimal strategy of always switching yields a success probability of \(2/3\), whereas quantum strategies employing entangled qutrit pairs and appropriate unitary operations can increase the winning probability to unity under ideal conditions. A QKD protocol based on this quantum Monty Hall game was proposed by Quezada \emph{et al.} \cite{quezada2020quantum}, and devising a fully device‐independent variant remains an open problem.

     \item \textit{RGB game:} Alice and Bob are the two players who need to provide an answer (a color). Albert is a friend of Alice and is located with her (on the moon in the thought experiment). Albert randomly picks one of the three colors (red, green, or blue) and tells Alice. This color is the one Alice is forbidden to answer with.

Boris is a friend of Bob and is located with him (on Earth). Boris independently and randomly picks one of the three colors and tells Bob. This color is the one Bob is forbidden to answer with.
Therefore, the "queries" in the RGB game refer to Albert telling Alice the color she cannot choose and Boris telling Bob the color he cannot choose. Alice and Bob must then choose one of the remaining two colors as their answer. They win if their chosen colors are different
     
     % In the RGB Game \cite{coiteux2019rgb}, two verifiers, Albert and Boris, interrogate two provers, Alice and Bob. The provers are far enough from each other that communication between them is impossible. Each prover may be independently queried one of three possible colours: Red, Green or Blue. To win the game, Alice and Bob answers should not match. The winning probability of the best classical strategy is $\frac{8}{9}\approx 0.89$, while the quantum strategy yields $\frac{11}{12} \approx 0.92$. This game has the potential to be used for QKD and DIQKD.
     % quantum key distribution protocols.
 \end{enumerate}

\begin{table*}[!h]
    \centering
    \caption{Loopholes in DIQKD. \label{loophole}}
    \begin{tabular}{| m{2cm} | m{5cm} | m{4cm}| m{5cm}|}
    \hline
    \textbf{Reference} & \textbf{Strategy} & \textbf{Locality Loophole} & \textbf{Detection Loophole} \\ \hline
    \end{tabular}
    \begin{supertabular}{| m{2cm} | m{5cm} | m{4cm}| m{5cm}|}
    \hline 
    \cite{mayers1998quantum} (1998) & Self-checking source & Ensures genuine entanglement & Mitigates inefficiencies via source verification \\[2ex]
    \hline
    \cite{kent2005causal} (2005) & Localized collapse process & State reduction events are space-like separated & N/A \\[2ex]
    \hline
    \cite{colbeck2009quantum} (2009) & Bell test for apparatus check & N/A & Verifies detector efficiency pre-key generation \\[2ex]
    \hline
    \cite{acin2007device} (2007) & Outcome substitution & N/A & Replaces non-clicks with noise \\[2ex]
    \hline
    \cite{gisin2010proposal} (2010) & Heralded qubit amplifier & N/A & Amplifies weak signals via heralding \\[2ex]
    \hline
    \cite{pitkanen2011efficient} (2011) & KLM-based heralding & N/A & Increases detection efficiency \\[2ex]
    \hline
    \cite{curty2011heralded} (2011) & Entanglement swapping relay & N/A & Overcomes channel losses \\[2ex]
    \hline
    \cite{vallone2013einstein} (2013) & Non-maximally entangled states (NMES) & N/A & Lowers threshold (66\% efficiency sufficient) \\[2ex]
    \hline
    \cite{brunner2013proposal} (2013) & Cavity QED heralding & Minimizes communication via fast spin readout & Near-unit efficiency spin measurements \\[2ex]
    \hline
    \cite{mattar2013device} (2013) & Heralded entanglement mapping & Sufficient spatial separation & Heralding confirms photon arrival \\[2ex]
    \hline
    \cite{vallone2014loss} (2014) & NMES protocol (B-92 variant) & N/A & Improves efficiency for 1sDI-QKD \\[2ex]
    \hline
    \cite{lim2013device} (2013) & Local Bell tests in lab & Local tests avoid channel loss issues & Insensitive to channel losses \\[2ex]
    \hline
    \cite{li2014quantum} (2014) & 2D Hilbert space encoding & N/A & Heralding minimizes loss, near-unit spin efficiency \\[2ex]
    \hline
    \cite{marshall2014device} (2014) & High-efficiency detectors, heralded mapping & Enforces sufficient separation & Robust mapping reduces detection issues \\[2ex]
    \hline
    \cite{hensen2016loophole} (2016) & Loophole-free Bell test with diamond spins & 1.3\,km separation closes locality loophole & N/A \\[2ex]
    \hline
    \cite{monteiro2017heralded} (2017) & Heralded photon amplification & N/A & Mitigates transmission losses \\[2ex]
    \hline
    \cite{zapatero2019long} (2019) & Photonic DIQKD with qubit amplifiers & Isolation of parties enforced & Assigns outcomes to close detection loophole \\[2ex]
    \hline
    \end{supertabular}
\end{table*}

\section{Games in DIQKD}
Nonlocal games provide a device-independent way to certify and harvest quantum correlations: by playing a Bell-type 'game' and observing a winning probability above the classical limit, Alice and Bob can generate a raw key and simultaneously bound any eavesdropper's knowledge, all without making assumptions about the inner workings of their measurement devices. In the DIQKD setting, a variety of games (CHSH, MPG etc.) have been adapted to test for nonlocality, extract secure bits, and tailor security proofs. The following subsections survey how specific games are implemented within DIQKD protocols.

\paragraph{DIQKD using CHSH game} 
In the CHSH game, as Alice and Bob choose their bases independently, there is a chance that they sometimes produce uncorrelated outcomes, which makes it difficult to produce reliable keys. To solve this problem, they require at least one choice of basis that gives them 
near perfect outcome. Therefore, we introduce an additional input choice for Bob as $y=2$. This input choice enables Bob to measure his qubits on a standard basis, which will result in near perfect outcome, as discussed in Section \ref{game1} (i.e. when two parties are entangled and measured in their standard basis, the chances of similar outcomes are near perfect). Hence, Alice's inputs become $x=\{0,1\}$ and for Bob's become $y=\{0,1,2\}$.  

A typical protocol for key generation based on the CHSH game proceeds as follows:
\begin{enumerate}
    \item Alice chooses an $n$ long string of bases uniformly at random $x = x_1\dots x_n \in \{0,1\}^n$, and her device measures the bases sequentially to generate a string $k_A=k_{1A},\dots k_{nA}$.
    \item Bob does the same; chooses his bases uniformly at random $y = y_1\dots y_n \in \{0,1,2\}^n$  and generate the string  $k_B=k_{1B},\dots k_{nB}$.
    \item Alice and Bob communicate the choices of their bases to each other through a classically authenticated channel (CA).
    \item Alice randomly selects a certain portion of indices of $Sub_A = k_A$, whose size is $n/2$ and announces $Sub_A$ to Bob. Bob then uses the $Sub_A$ to further divide $Sub_A$ into two parts: $Sub_{hd}$ and $Sub_{sd}$ (subscripts $hd$ stands for Hadamard basis and $sd$ for Standard basis).
    \item Bob aligns his outcomes of $y=2$ with $Sub_{sd}$ and the outcomes of $y=\{0,1\}$ with $Sub_{hd}$.
    \item Alice and Bob publicly reveal and compare the outcome strings corresponding to the indices in \(Sub_A\).
    \item Using the outcomes for \(Sub_{hd}\) they compute the CHSH winning probability \(p_{win}^{CHSH}\); simultaneously, they use the outcomes for \(Sub_{sd}\) to calculate the near-perfect correlation probability \(p_{match}\). 
    \item If either \[p_{win}^{CHSH} < \cos^2\left(\frac{\pi}{8}\right) \quad \text{or} \quad p_{match} < \frac{1}{2},\] they abort the protocol; otherwise, they proceed.
    \item The final key is generated from the remaining indices (denoted as \(Rem_A\)) that were not used in the tests, yielding the secure keys \(k_A^{Rem_A}\) and \(k_B^{Rem_A}\) for communication. 
\end{enumerate}

Ultimately, to ensure the secrecy and correctness of their final keys, Alice and Bob perform two crucial post‐processing steps. First, they apply information reconciliation, a procedure that corrects any discrepancies between their key strings. Next, they employ privacy amplification, which shortens the reconciled key to eliminate any partial information that an adversary might have obtained.

There has been remarkable research related to DIQKD using CHSH games. Acin \emph{et al.}  \cite{acin2007device} first initiated the idea of DIQKD when they upgraded their own QKD protocol \cite{acin2006efficient} to device-independence. Later, Pironio \cite{pironio2009device} extended Acin's work and gave the security proof of the DIQKD protocol proposed by Acin \emph{et al.}  \cite{acin2007device}. Despite the obvious security benefits, two significant loopholes limited the wide adoption of DIQKD. The first loophole is the locality loophole, which is the possibility of particles or devices communicating with each other during the protocol. This loophole can easily hamper the entanglement test (CHSH game) performed before the key generation process. The second loophole is called the detection loophole, which is the possibility of a few particles going undetected, yielding biased results for entanglement tests. Researchers have been actively investigating ways to address these loopholes; Mayers and Yao \cite{mayers1998quantum} gave the concept of self-checking sources using nonlocal correlation. Similarly, Hensen \emph{et al.} \cite{hensen2016loophole} have experimentally closed the locality loophole by separating the electron spin in an artificial diamond and the detector 1.3 km apart, making communication impossible. The related work on loopholes is summarized in Table \ref{loophole}.

\paragraph{DIQKD using MPG game} The quantum strategy of the MPG game confirms the correlation of the outputs for their respective inputs. The entanglement confirmation gives us the liberty to use it in Quantum key distributions, especially in DIQKD. To make our results binary, we identify the eigenvalues 1 and -1 as 0 and 1, respectively. The protocol is similar to the one used for CHSH-based DIQKD, with the exception of the entanglement confirmation step. The protocol is as follows:
\begin{enumerate}
    \item Alice and Bob are given the input $x_n$, $y_n$ $\in \{0,1,2\}$ from a third party, where $n$ being the number of rounds they play. They feed their devices with the inputs and get the output bits based on the equations $[A_0^{x_{n}},A_1^{x_{n}},A_2^{x_{n}} = A_0^{x_{n}} \oplus A_1^{x_{n}}]$ for Alice and $[B_0^{y_{n}},B_1^{y_{n}},B_2^{y_{n}} = B_0^{y_{n}} \oplus B_1^{y_{n}}\oplus 1]$ for Bob.
    \item The inputs $x_n$, $y_n$ are treated as public keys and are announced. However, the outputs $A_{y_{n}}^{x_{n}}$ and $B_{x_{n}}^{y_{n}}$ are considered private keys.
    \item Alice randomly chooses some part of the index of the total number of rounds performed $N_A$ and communicates it to Bob. They use bits according to the index $N_A$ provided by Alice as raw keys \textbf{a} and \textbf{b}. The bits are used to play MPG game. If the average winning probability falls below the expected threshold ($\omega < \omega_{exp}$), the protocol is aborted; otherwise, it proceeds.
    \item They apply data reconciliation protocols on the raw keys, including error correction and privacy amplification, to obtain the final secure keys $\kappa_A$ and $\kappa_B$.
\end{enumerate}
This DIQKD protocol, which involves the MPG game, is presented in Zhen \emph{et al.} \cite{zhen2023device}. In addition to the protocol, they also provided the security analysis, comparing its key rates with the biased CHSH game \cite{lawson2010biased}. The results show that the MPG-DIQKD surpasses the CHSH-DIQKD in key rates if the optimal quantum strategy is faithfully implemented.

In summary, while the CHSH game currently remains the most practical for DIQKD implementations due to its robustness under realistic conditions, these alternative non-local games (MPG, GHZ, Monty Hall, and RGB) offer intriguing possibilities for expanding DIQKD protocols. However, their reliance on ideal conditions poses significant challenges for practical implementation.
%%%%%%%%%%%%%%%%%%%%%%%%%%%%%%%%%%%%%%%%%%%%%%%%%%%%%%%%%

\section{Security of DIQKD} \label{security}
The security in DIQKD is based on the violation of Bell's inequality, which certifies that Alice and Bob's devices are performing quantum measurements on entangled states. Through this violation, they can bound the information that any adversary (even one controlling the devices) has about the shared key. There are two main steps to prove security in DIQKD. 
\begin{enumerate}
    \item determining the lower bound on the conditional min-entropy in every single round of the protocol.
    \item combining these bounds into a collective bound for the combined key.
\end{enumerate}

To evaluate the security of the system, these steps are performed in various situations in which Eve is given some power, called \emph{attacks}.

In QKD, there are three primary strategies that an eavesdropper, Eve, might use to intercept communications: individual attacks, collective attacks, and coherent attacks. These classifications emphasize the varying levels of complexity and the potential impact of different eavesdropping techniques.

\subsection{Individual Attacks}
 Individual attacks involve an eavesdropper (commonly referred to as Eve) intercepting and measuring each quantum signal independently, without retaining quantum memory of past interactions. This means that Eve's measurements on each quantum state are separate and do not depend on measurements of other states. Such attacks are generally simpler to implement, but may be less effective than more complex strategies \cite{murtalecture}.

 A sequential attack \cite{roy2024sequential} is a specialized form of individual attack in which Eve performs a series of unsharp (weak) measurements on the quantum states as they travel from the sender (Alice) to the receiver (Bob). In this strategy, Eve aims to extract information about the key while introducing minimal disturbances to the quantum states, thereby preserving the observable quantum correlations that Alice and Bob rely upon for security. This subtle interference allows Eve to gain partial knowledge of the key without being easily detected. Sequential attacks exploit the delicate balance between measurement disturbance and information gain inherent in quantum mechanics. By carefully tuning the parameters of her unsharp measurements, Eve can maximize her information acquisition while minimizing the risk of detection. This makes sequential attacks particularly insidious, as they can significantly compromise the security of QKD protocols if not properly accounted for. 

 Understanding and mitigating individual attacks, especially sequential attacks, is crucial for the robust implementation of QKD systems. Developing countermeasures against such eavesdropping strategies ensures the integrity and confidentiality of the cryptographic keys generated through quantum communication.
\subsection{Collective attacks}
In these attacks, Eve interacts with each individual signal (qubit) transmitted between Alice and Bob independently. However, instead of measuring the information right away, Eve stores the outcomes in a quantum memory and waits until Alice and Bob exchange classical information (such as their basis choices), then performs measurements to extract information about the shared key. The way to quantify the statistical significance of the observed data, is to apply the Chernoff bound, which will provide a way to differentiate between normal fluctuations and actual attacks. Alice and Bob utilize the Chernoff bound to determine the maximum probability that the observed Bell inequality violation could occur due to statistical fluctuations, even in the absence of any eavesdropping by Eve. This calculation helps them assess whether the observed violation is significant enough to indicate potential eavesdropping or if it can be attributed to random chance. 
If the observed violation is too low with respect to this probability, they can conclude that Eve might be interfering and decide to abort the protocol. The Chernoff bound is given by:
\begin{align}
    Pr\Big(\Big | \frac{1}{n} \sum_{i=1}^n X_i - \mu \Big| > \alpha\mu \Big) \leq 2 e^{-\frac{\alpha^2 \mu n}{3}}
\end{align}
where $X_i$ is the i.i.d. random variable taking values in \{0,1\}, $\mu$ is the expectation value of $X_i$, and $\alpha$ is the scaling factor that describes how much the actual sum of the random variables is below its expected value. The probability that results in giving the direct bound on the guessing probability of the device is $p_{ind} \leq 1/2 + 2(2\delta)^{1/2}$ ($\delta$ is the deviation between the ideal and actual winning probability), using this probability, we can obtain the bounded min-Entropy as: 
\begin{align}
    H_{min}(X_j|E) \geq 1 - C\sqrt{\delta} \label{min}
\end{align}
where $C$ is a small constant. As our assumption for this attack is that the device behaves identically and independently in every round, the min-entropy bound in Eq. (\ref{min}) can also be applied to the rounds designated to raw keys in addition to the rounds specified for the testing of the CHSH game. This feature allows us to add the entropies, which yields the final bounded min-entropy as:
\begin{align}
    H_{min}(X_R|E) \geq |R|(1 - C\sqrt{\delta}) \label{min2}
\end{align}
where $|R|$ is the number of rounds designated for raw keys. Considering that privacy amplification and information reconciliation are performed correctly, Alice and Bob can generate secure keys.

The main idea of device-independent collective attacks in general cryptographic systems was proposed by Acin \emph{et al.} \cite{acin2007device}, but the actual discussion of collective attacks on device-independent QKD was initiated by Pironio \emph{et al.} \cite{pironio2009device}. After the initial work, a series of research followed (listed in Table \ref{secure_new}). SDI-QKD initially was the trusted DIQKD system that was considered effective against collective attacks. However, in recent years, we have observed that most of the work on collective attacks is done using MDI-QKD. Looking at MDI-QKD, we observe that MDI-QKD is often realized as a Continuous-Variable system. This means that Continuous-Variable MDI-QKD is more efficient than its DV counterpart against collective attacks \cite{pirandola2015reply}.

\begin{table*}[!h]
    \centering
    \caption{Comparison of DIQKD Systems Secure Against Collective Attacks}
    \begin{tabular}{|l|l|l|p{11cm}|}
    \hline
    \textbf{Ref}  &  \textbf{Platform} &  \textbf{Protocol Type} &  \textbf{Key Contributions/Comments} \\ \hline
    \cite{acin2007device} (2007) & DV & DIQKD & Introduced an optimal collective attack model via a tight bound on the Holevo information between one party and the eavesdropper. \\ \hline
    \cite{pironio2009device} (2009) & DV & DIQKD & Provided a detailed security proof for DIQKD against collective attacks, assuming independent and identical actions in each round. \\ \hline
    \cite{woodhead2012semi}  (2012) & DV & SDI-QKD & Proposed a prepare-and-measure scheme based on BB84 and CHSH-type estimation, with security proven under collective attacks assuming a bound on the Hilbert space dimension. \\ \hline
    \cite{wang2014security}  (2014) & DV & SDI-QKD & Developed a practical SDI-QKD protocol using four preparation states and three measurement bases, with security analyzed via min-entropy and dimension witnesses. \\ \hline
    \cite{tan2016biased} (2016) & DV & DIQKD & Proposed a biased random number generation protocol based on Bell's theorem, demonstrating secure bit string bias under collective attacks. \\ \hline
    \cite{woodhead2016semi} (2016) & DV & SDI-QKD & Derived an analytic lower bound on the asymptotic secret key rate for an entanglement-based BB84 variant using unknown qubit POVMs. \\ \hline
    \cite{yang2016measurement}  (2016) & DV & MDI-QKD & Demonstrated that MDI-QKD based on entanglement can be made loss-tolerant, with a tight bound on Eve’s Holevo information under collective attacks. \\ \hline
    \cite{wang2018self}  (2018) & CV & MDI-QKD & Proposed a self-referenced CV-MDI-QKD scheme that removes the need for transmitting a high-brightness local oscillator; security simulations indicate only a slight reduction in transmission distance under collective attacks. \\ \hline
    \cite{liao2018dual}  (2018) & CV & MDI-QKD & Introduced a dual-phase-modulated plug-and-play CV-MDI-QKD protocol and derived security bounds against optimal Gaussian collective attacks. \\ \hline
    \cite{ma2019long}  (2019) & CV & MDI-QKD & Developed a long-distance CV-MDI-QKD protocol with discrete modulation; security analysis confirms robustness against arbitrary collective attacks using decoy states. \\ \hline
    \cite{kang2019measurement}  (2019) & DV & MDI-QKD & Proposed an MDI-QKD protocol with inaccurate coherent sources; provided security proofs against collective attacks with experimental simulation of the security bound. \\ \hline
    \cite{wu2019security} (2019) & CV & MDI-QKD & Performed security analysis for a passive CV-MDI-QKD protocol with minimal public communication, deriving asymptotic security bounds under collective attacks. \\ \hline
    \cite{ma2019security} (2019) & CV & MDI-QKD & Addressed the security of CV-MDI-QKD with imperfect state preparation over lossy and noisy channels, deriving a lower bound on the secret key rate under arbitrary collective attacks. \\ \hline
    \cite{tan2020advantage} (2020) & DV & DIQKD & Investigated advantage distillation using two-way communication in DIQKD to improve noise tolerance, and provided verifiable conditions for security against collective attacks. \\ \hline
    \cite{datta2022device} (2022) & DV & DIQKD & Introduced an approach where an optimal Bell inequality is postselected from measurement data, with finite-size key analysis demonstrating security against collective attacks. \\ \hline
    \cite{zhou2023plug}  (2023) & CV & MDI-QKD & Proposed a plug-and-play CV-MDI-QKD protocol using Gaussian-modulated coherent states, with security bounds analyzed under Gaussian collective attacks. \\ \hline
    \end{tabular}
    \vspace{0.2cm}
    \label{secure_new}
\end{table*}

\subsection{Coherent Attacks}
Coherent attacks can be considered advanced variants of collective attacks, where Eve uses more powerful strategies. In these attacks, Eve can create a global quantum state that links multiple rounds together, and her attacks can be dependent on previous rounds, which makes security analysis much more complex. 
The complexity arises because Eve’s devices can have quantum memory, allowing her to store and process information in multiple rounds. This inter-round correlation means that even if Alice and Bob perform statistical checks (like Bell tests) on a random subset of rounds, they cannot reliably extrapolate the security of the remaining rounds. The devices may behave honestly during the test rounds, but maliciously during key generation rounds.
% In particular, in coherent attacks, Eve's devices can have memory, which means that their behavior in one round can depend on what happened in previous rounds. This makes it difficult to predict their behavior simply by looking at the testing rounds.
Consequently, we cannot directly apply results from the testing rounds to the rounds used for the raw key. To address this problem (where devices may have memories), a special Martinangle inequality known as \textit{Azuma inequality} is used. By applying Azuma inequality, one can bound the min-entropy per round even when the devices have memory. This allows us to estimate the uncertainty (min-entropy) for each key bit, despite the rounds not being independent. Consequently, we can obtain a bound on the min-entropy for the raw key rounds.

Another challenge with coherent attacks is that in these attacks, Eve can create a global entangled state that involves multiple rounds. This means that we can no longer simply add the min-entropy from each round to get the total min-entropy (which was the case with collective attacks). To address this problem, the Entropy Accumulation Theorem (EAT) \cite{dupuis2020entropy} is used. EAT helps when the process generating the outcomes (such as Alice and Bob's key bits) happens sequentially. Mathematically, if each round $j$ produces at least $h_j$ bits of min-entropy the EAT guarantees that the total min-entropy $H_{min}$ for $n$ rounds is at least the sum of the individual min-entropies:
\begin{align}
    H_{min}(X_R|E) \geq \sum_{j\in R} h_j
\end{align}
where $R$ represents the rounds used for the raw key. There is a small loss in the security parameter, $\epsilon$, which measures how closely the total min-entropy approximates the ideal case. This small loss comes from the fact that we are dealing with potentially dependent rounds, but the overall security is still sufficiently high. %Once its applied, EAT serves as a powerful tool,  it provides almost the same results that we were able to derive in the case of collective attacks.

Another important aspect of analyzing coherent attacks involves estimating the total uncertainty, or entropy, of the raw key generated by Alice and Bob. While the Entropy Accumulation Theorem allows us to bound the total min-entropy by summing per-round contributions, an even more powerful result emerges in the asymptotic regime. When the number of rounds is large, and the quantum states across rounds are identically and independently distributed (i.i.d.), the total smooth min-entropy becomes approximately equal to the von Neumann entropy. This relationship is formalized by the Asymptotic Equipartition Property (AEP) \cite{tomamichel2009fully}, which states that the smooth min-entropy converges to the von Neumann entropy as the number of rounds increases. Since the von Neumann entropy generally provides a tighter (i.e., higher) bound than the min-entropy, using AEP allows for stronger security estimates. Consequently, this leads to more accurate evaluations of the achievable key rate.

Table \ref{coherent} summarizes the main published results related to coherent attacks in DIQKD. The first work to recognize coherent attacks in DIQKD was introduced by McKague \cite{mckague2009device}, who extended the security analysis of collective attacks provided in \cite{pironio2009device} to coherent attacks. However, compared to collective attacks, research on coherent attacks is limited as these attacks are very difficult to realize. However, researchers have analyzed the security against coherent attacks in 1sDI-QKD, MDI-QKD, and FDI-QKD, as detailed in Table \ref{coherent}.

\begin{table*}[!h]
\centering
\label{coherent}
\caption{Comparison of DIQKD Systems Secure Against Coherent Attacks}
\begin{tabular}{|l|l|l|p{11cm}|}
\hline
\textbf{Ref} & \textbf{Platform} & \textbf{Protocol Type} & \textbf{Key Contributions/Comments} \\ \hline
\cite{wu2016continuous} (2016) & CV & MDI-QKD & Analyzed security of a CV MDI protocol against both entangling cloner and coherent attacks, showing that coherent attacks significantly reduce the key rate. \\ \hline
\cite{broadbent2015device}  (2015) & DV & DIQKD & Utilized generalized two-mode Schrödinger cat states; proved security against collective attacks and, in certain cases, coherent attacks. \\ \hline
\cite{marshall2014device} (2014) & CV & DIQKD & Proposed a CV DIQKD protocol based on the Gottesman-Kitaev-Preskill encoding; security is derived from DV techniques for collective attacks and, under memoryless assumptions, coherent attacks. \\ \hline
\cite{vazirani2019fully}  (2019) & DV & DIQKD & Established full device-independent security of a variant of Ekert's protocol against general coherent attacks by assuming only quantum mechanics and spatial isolation. \\ \hline
\cite{mckague2009device} (2009) & DV & DIQKD & Extended earlier collective attack security proofs to coherent attacks, assuming measurement devices are memoryless. \\ \hline
\cite{masini2024one} (2024) & DV & 1s-DIQKD & Demonstrated one-sided DIQKD secure against coherent attacks over long distances, requiring detection efficiencies above 50.1\% on the untrusted side. \\ \hline
\cite{branciard2012one}  (2012) & DV & 1s-DIQKD & Derived a security bound against coherent attacks under memoryless device assumptions, linking security with quantum steering. \\ \hline
\end{tabular}
\label{coherent_new}
\end{table*}

%%%%%%%%%%%%%%%%%%%%%%%%%%%%%%%%%%%%%%%%%%%%%%%%%%%%%%%%%

\section{Implementation of DIQKD}

Practically implementing the ideal case of DIQKD, known as Fully Device-Independent QKD (FDI-QKD), requires conducting a loophole-free Bell test over a significant distance while maintaining a low QBER. Given the limitations of the current infrastructure, achieving these stringent requirements poses a significant challenge.
Several attempts have been made to practically implement DIQKD on different experimental platforms, including photonic setups, atomic systems, and superconducting circuits. These experimental realizations are typically designed to demonstrate a violation of the CHSH inequality, as expressed in Eq. (\ref{S}). The degree of violation achieved varies depending on the experimental setup and the distance separating the communicating parties. Some recent experimental demonstrations of loophole-free Bell tests using various platforms are summarized in Table \ref{violations}.

As shown in Table \ref{violations}, photonic setups have been extensively studied among these platforms, making it the most mature approach for DIQKD experiments. Beyond photonics, matter-based systems such as trapped ions, nitrogen-vacancy (NV) centers in diamond, and superconducting qubits, have also been explored, offering unique advantages and challenges.
 
% Beyond photonics, matter-based systems 
\begin{table*}[t]
\centering
\label{violations}
\caption{Loophole-free Bell Violations Used for DIQKD Experiments by Physical Platform}
\label{tab:diqkd-experiments-updated}
\begin{tabular}{|p{2cm}| p{2cm} | p{1.5cm} | p{2.5cm}| p{7cm}|}
\hline
\textbf{Reference} & \textbf{Platform} & \textbf{Bell Test} & \textbf{Distance} & \textbf{Outcome} \\
\hline
\cite{Giustina2015} (2015) & Photonic  & CHSH & Short-to-moderate & First loophole-free Bell tests using photon entanglement and superconducting nanowire detectors. \\
\hline
\cite{big2018challenging} (2018) & Photonic  & CHSH & Short-to-moderate & Human-generated randomness was used to close the freedom-of-choice loophole. \\
\hline
\cite{liu2022toward} (2022) & Photonic  & CHSH & Short-to-moderate & Demonstrated high-fidelity entanglement distribution with improved key rates under loophole-free conditions. \\
\hline
\cite{rosenfeld2017event} (2017) & Atomic  & CHSH & $\sim$1.3 km & Achieved single-atom entanglement with event-ready detection in a loophole-free test. \\
\hline
\cite{nadlinger2022experimental} (2022) & Atomic  & CHSH & Moderate & Performed an ion-trap-based Bell test with enhanced coherence for DIQKD protocols. \\
\hline
\cite{zhang2022device} (2022) & Atomic  & CHSH & Moderate-to-long & Explored entanglement swapping between separate atomic ensembles for scalable DIQKD. \\
\hline
\cite{Hensen2015loophole} (2015) & Solid-State  & CHSH & 1.3 km & NV center spins in diamond produced a loophole-free Bell test, closing both detection and locality loopholes. \\
\hline
\cite{Storz2023Loophole} (2023) & Solid-State  & CHSH & Lab scale; chip-based & Demonstrated a loophole-free Bell test with on-chip superconducting qubits—a key step toward integrated DIQKD. \\
\hline
\end{tabular}
\end{table*}

\subsection{DIQKD Based on Photonic Systems}
Photonic systems are the most widely explored platforms for DIQKD, largely due to the relative ease of generating and measuring entangled photons in various degrees of freedom, such as polarization, time-bin, or orbital angular momentum (OAM). 
Most photonic DIQKD demonstrations rely on nonlinear processes such as spontaneous parametric down-conversion \cite{zapatero2023advances} or spontaneous four-wave mixing \cite{liu2024reconfigurable} to produce entangled photon pairs, enabling high rates of entanglement generation. A central requirement is performing a loophole-free Bell test, which necessitates addressing the detection loophole through high-efficiency single-photon detectors, the locality loophole via space-like separation of measurement stations, and the freedom-of-choice loophole through unbiased and fast random number generators. Although no experiment has fully realized large-scale DIQKD, many photonic systems feature preliminary key distillation steps applying error correction and privacy amplification to data that violate the Bell inequality under partial device-independence assumptions. Current challenges for implementing DIQKD on photonics platforms include mitigating loss, extending transmission distance in optical fibers or free-space links, ensuring synchronization and phase coherence in multi-photon interference setups, and scaling beyond laboratory conditions.

A notable example is the work of Liu \emph{et al.} \cite{liu2022toward}, who experimentally demonstrated a CHSH game with Bell's violation across a 20-meter fiber link. Their setup achieved a CHSH winning probability of approximately 0.7559, corresponding to a Bell-CHSH parameter $S=2.0472$, this result surpasses the classical limit of $S=2$ and exemplifies how photonic platforms can reliably distribute entanglement over modest distances for DIQKD-related tests.

Recent advances in integrated photonics and efficient single-photon sources, along with improved stability and lower losses, promise to bring photonic DIQKD closer to practical implementation, potentially making it a foundational technology for future secure quantum communication networks.
  
\subsection{DIQKD based on Matter-Based systems}

Matter-based systems provide a promising alternative to photonic platforms for implementing DIQKD, primarily because of their excellent coherence properties and the precise control that can be exerted over internal atomic states. These experiments prepare neutral atoms or trapped ions in entangled configurations, often mediated by photon exchange or direct interactions in carefully controlled environments such as ion traps or optical lattices. A key feature of matter-based setups is their long coherence times, which allow entanglement to persist over comparatively extended periods, facilitating more reliable Bell tests. However, to fully leverage these advantages, experiments must overcome fundamental loopholes. High-fidelity state preparation and efficient measurements are crucial to detect loopholes. Additionally, the locality loophole requires strict spatial or temporal separation of atomic ensembles, while the freedom-of-choice loophole necessitates rapid and unbiased switching of measurement bases.
% high-fidelity state preparation and measurement are crucial to mitigating the detection loophole, while strict spatial or temporal separation of atomic ensembles and rapid switching of measurement bases help close the locality and freedom-of-choice loopholes.

Recent experiments underscore the steady progress being made. For example, Nadlinger \emph{et al.} \cite{nadlinger2022experimental} demonstrated a trapped-ion-based heralded entanglement system achieving a CHSH value of $S\approx2.64$ and QBER of about $1.8\% $ across a 2m ion-trap link, illustrating the practicality of DIQKD methods over short laboratory distances. Meanwhile, Zhang \emph{et al.} \cite{zhang2022device} reported an event-ready entanglement setup using two independently trapped single rubidium atoms separated by 400m. Their system reached an entanglement fidelity of $F\geq0.892$, with a significant Bell violation $S=2.578$ and a remarkably low QBER of $0.078$. The authors implemented a DIQKD protocol that achieved a secret key rate of 0.07 bits per entanglement generation event in the asymptotic limit, indicating the potential for device-independent communication in real-world scenarios.

Despite significant progress, fully loophole-free demonstrations of large-scale DIQKD remain challenging in matter-based systems, partly due to the complexity of maintaining entanglement over long distances.

\subsection{Full DIQKD Demonstration}
 Conducting a loophole-free Bell test at high rates and over long distances is challenging. Scaling up these experiments to generate and distribute secret keys with low error rates and robust privacy amplification adds another layer of difficulty. Even minor imperfections in sources, detectors, or random number generators can open side channels, undermining the device-independent security guarantee. DIQKD often assumes perfectly isolated laboratories and no hidden radio transmitters and classical side channels, which are demanding conditions in practical setups.

 Although milestones have been achieved in loophole-free Bell tests and partial or proof-of-concept DIQKD experiments, a fully implemented DIQKD protocol (from entanglement generation and Bell violation through key distillation under complete device-independence) has yet to be experimentally realized. Current research focuses on bridging these remaining gaps to bring DIQKD from the lab into practical use.
 
\section{Open Problems} \label{open}
Despite the significant progress made so far in DIQKD, several key challenges remain unsolved, particularly in developing new protocols that are more robust against coherent attacks and designing innovative quantum games.

\paragraph{Protocols.} One one the most challenging (and interesting) challenges in DIQKD is the development of new DIQKD protocols that maintain high key rates while providing robust security against these complex adversarial strategies. Achieving this goal will likely require new mathematical techniques for handling correlated rounds, as well as further exploration of the conditions under which EAT can be most effectively applied.

\paragraph{Quantum Games.} The design of quantum games, such as the CHSH game, is central to certifying the security of DIQKD protocols. However, current games may not fully capture the range of quantum strategies an adversary could use in a real-world scenario, particularly in the presence of noise or device imperfections. There is a growing need for new quantum games that can offer better security guarantees, especially under coherent attacks or in more practical, noisy environments.
These games should be designed to maximize quantum advantage and be easily testable with real-world quantum devices. The challenge lies in finding games that balance theoretical rigor and experimental feasibility, offering new avenues to certify security in DIQKD.

\paragraph{Satellite-based Communication.} Another promising direction is the integration of DIQKD protocols into satellite-based quantum communication networks, enabling global coverage and bypassing the limitations of ground-based fiber networks. However, achieving device independence in a space environment involves numerous open challenges, including coping with atmospheric losses, beam wander, and stringent payload constraints. 

\paragraph{On-chip Design}. Another avenue is the development of on-chip or integrated photonic solutions, which promise scalability and robustness by reducing the size and complexity of DIQKD systems. Integrated photonic platforms would benefit from mature semiconductor fabrication processes, potentially allowing higher repetition rates and improved stability. These chip-based solutions, combined with new methods for distributing entanglement over metropolitan or intercontinental distances, could pave the way for secure, low-cost quantum networks. 

\paragraph{Engineering.} Overcoming engineering hurdles, such as miniaturizing high-efficiency detectors and maintaining low-loss waveguides at telecom wavelengths, remains an open problem.

\section{Conclusion} \label{conclusion}
In this survey, we have explored the landscape of Device-Independent Quantum Key Distribution (DIQKD), a promising approach to secure quantum communication that guarantees security even with untrusted devices. By leveraging the violation of Bell's inequalities, DIQKD provides a level of security that is fundamentally different from traditional QKD protocols.
We examined different types of DIQKD protocols, each offering distinct advantages based on different assumptions and settings. These protocols highlight the adaptability of DIQKD across various practical scenarios, though each comes with unique challenges in terms of implementation and security guarantees.
A crucial element of DIQKD lies in using quantum games, such as the CHSH game, which serve as a foundation for certifying security. These games provide a way to test quantum correlations and ensure that the devices behave as expected, even under adversarial conditions. However, as discussed, there is a growing need for the development of new quantum games to address the limitations of current protocols, particularly in noisy or imperfect environments.
Finally, DIQKD security remains a significant open research area, especially in the face of advanced attacks such as collective and coherent attacks. We explored the tools and techniques used to ensure security, including the Entropy Accumulation Theorem (EAT), which has provided a pathway for analyzing security under more complex adversarial models. However, designing practical and robust protocols against these powerful attacks remains an open problem.

%%%%%%%%%%%%%%%%%%%%%%%%%%%%%%

%%%%%%%%%%%%%%%%%%%%%%%%%%%%%%%%%%%%%%%%%%%%%%
%%%%%% SECCIONES DE DISEÑO Y DESARROLLO %%%%%%
%%%%%%%%%%%%%%%%%%%%%%%%%%%%%%%%%%%%%%%%%%%%%%
%%%%%%%%%%%%%%%%%%%%%%%%%%
\bibliographystyle{ieeetr}
\bibliography{SurveyDIQKD}
\end{document}